\documentclass[preprintnumbers, prd, showpacs, twocolumn, floatfix, preprintnumbers, letterpaper, amsmath, amssymb, superscriptaddress]{revtex4}

\usepackage{amssymb, amsmath, bm, dcolumn, epsf, graphicx, latexsym, mathbbol, slashed, amsthm, bm, mathrsfs, float, subcaption}
\usepackage[colorlinks,linkcolor=red,urlcolor=blue,citecolor=blue]{hyperref}
\usepackage{tabularx}
\usepackage{booktabs}
\usepackage{comment}

\begin{document}

\newcommand{\dd}[0]{\mathrm{d}}

\title{Rigorous calculation of scalar scattering in Schwarzschild background: the convergence of partial-wave series and Poisson spot}
\author{Zhao Li}
\email{lz111301@pku.edu.cn}
\affiliation{Department of Astronomy, Peking University, Beijing, 100871, China}
\affiliation{Department of Astronomy, University of Science and Technology of China, Hefei, Anhui 230026, China, and\\ School of Astronomy and Space Science, University of Science and Technology of China, Hefei, Anhui 230026, China}
\author{Wen Zhao}
\email{wzhao7@ustc.edu.cn}
\affiliation{Department of Astronomy, University of Science and Technology of China, Hefei, Anhui 230026, China, and\\ School of Astronomy and Space Science, University of Science and Technology of China, Hefei, Anhui 230026, China}
\affiliation{College of Physics, Guizhou University, Guiyang, Guizhou 550025, China}

\date{\today}

\begin{abstract}
Black hole (BH) perturbation theory and the scattering models provide a powerful framework for studying gravitational lensing at the wave-optics level. However, conventional calculations encountered two issues: the divergence of the partial-wave series and the divergence of the Poisson spot near the optical axis. These issues hinder the accurate calculation of lensed waveforms and the study of polarization and wave characteristics in the lensing process, especially near the optical axis. This work demonstrates that both divergences stem from the asymptotic expansion of the radial wave function. By computing the scattered wave function at finite radii and avoiding the asymptotic expansion, we naturally obtain convergent results. We compute scalar waves scattered by (1) a weak-gravity body with Newtonian potential and (2) a Schwarzschild BH with Regge-Wheeler potential. In both cases, we analyze the convergence of the partial-wave series and present finite-luminosity diffraction patterns, with a bright Poisson spot. The above calculations are compared with the Kirchhoff diffraction integral in the near-axis regions and give consistent results. Our investigations provide a foundation for studying gravitational wave scattering by BHs and understanding lensing at the wave-optics level.
\end{abstract}

\maketitle

\section{\label{sec:intro}Introduction}

Gravitational lensing, a cornerstone prediction of general relativity \cite{weinberg,mtw,Wambsganss_1998,Bartelmann_2010,Dodelson_2017}, serves as a powerful probe for exploring cosmic matter distributions \cite{Refregier_2003,Cao_2012} and testing gravitational theories \cite{Liu_2022,Wang_2024}. In the era of multi-messenger astronomy, gravitational wave (GW) lensing has attracted extensive attention \cite{Ezquiaga_2021,Takahashi_2003,Meena_2019,Nakamura_1998}. While wide searches have been conducted for lensed GW signals in current observational data, significant evidence remains absent \cite{Haris_2018,Hannuksela_2019,McIsaac_2020,XiaoshuLiu_2021,LIGO_2021,Lo_2023,LIGO_2023}. Upcoming next-generation ground-based GW detectors are expected to observe millions of GW signals, with approximately 0.1–1\% expected to undergo strong lensing by the galaxies or clusters, producing detectable imprints in the GW signals \cite{Yang_2021,Lo_2025,Piorkowska_2013,Lin_2023}. 

The successful GW lensing detections critically depend on the accuracy of waveform templates. Three main distinct theoretical approaches to GW lensing have been applied. The first approach focuses on the lowest-order approximation of GWs, \emph{geometric optics}. Similar to the electromagnetic waves, GWs correspond to an undiscovered massless spin-2 gauge boson, referred to as gravitons,  propagating along the null geodesics on curved spacetimes \cite{Maggiore_2008}. When encountering massive objects, their trajectories are deflected by the external gravitational field, forming multiple images on the lens plane. For specified mass distributions, the deflection angle and time delay can be derived by integrating the geodesic equations \cite{Bozza_2010,Keeton_2005,sereno_2006}. Combining the lens equation and Jacobian matrix, this approach yields the magnification of GW amplitudes \cite{Keeton_2005,sereno_2006,Li_2025} and relates the lensed and unlensed signals through a frequency-independent transmission factor. However, this approach completely ignores the wave effects such as interference and diffraction, and fails to capture the frequency dependence of lensing phenomena, resulting in a series of degeneracies between: magnification and luminosity distance, time delay and initial phase, gravitational Faraday rotation and initial polarization angle, leaving great difficulties for practical detection \cite{Ezquiaga_2021,Li_2022}.

Beyond geometric optics and building upon Huygens' principle, \emph{Kirchhoff diffraction theory} formulates the transmission factor as an integral over all possible GW paths from the wave source, through the lens plane, to the observer \cite{Baraldo_1999,Born_2013,Guo_2020}. This framework offers three key advantages: computational efficiency, adaptability to various lens models \cite{Keeton_2002}, and accurate characterization of wave-optical phenomena, including interference and diffraction \cite{Takahashi_2003,Dai_2017,Lo_2025,Yuan_2025}. As such, it has emerged as the predominant method for constructing lensed waveform templates, effectively bridging the gap between geometric and wave optics. However, this approach remains fundamentally based on the short-wavelength approximation of the linearized Einstein field equations (LEFE), and this treatment neglects the intrinsic spin-2 nature of GWs, approximating their amplitude as a scalar field \cite{mtw,Cusin_2020,Dalang_2022}. These constraints underscore the need to develop a comprehensive, full-wave theory for a deeper understanding of gravitational lensing phenomena.

The most fundamental wave-optics approach is to directly solve the LEFE in curved backgrounds as done by Refs.\,\cite{He_2021,He_2022,Yin_2024}, which remains a challenging and computationally expensive mission. Fortunately, the black hole (BH) perturbation theory, pioneered by Regge and Wheeler \cite{regge_wheeler_1957} and subsequently developed by Zerilli, Teukolsky, and others \cite{zerilli_1970a,zerilli_1970b,bardeen_1973,teukolsky_1973,press_1973,teukolsky_1974,chandrasekhar_1978a,chandrasekhar_1978b,sasaki_nakamura_1982}, provides an analytical framework to describe the GW propagation in the BH backgrounds. Established on this formalism, the early contributions to BH scattering include \cite{Hildreth_1964,Matzner_1968,Chrzanowski_1976,Fabbri_1975,Sanchez_1976,Sanchez_1978a,Sanchez_1978b,Handler_Matzner_1980}, which concentrate on calculating the differential cross section (DCS) and absorption cross section. These results are systemically extended by Refs.\,\cite{Andersson_1995,Glampedakis_2001,Dolan_2008a,Dolan_2008b}. Serving GW observation, recent works have moved to calculate the scattered gravitational waveform \cite{Bao_2022,Pijnenburg_2024b,Chan_2025}. This full-wave approach fundamentally captures all essential characteristics of scattered GWs, including wave interference, time delay, and gravitational Faraday rotation. 

However, practical computation of this scattering model via the partial-wave method encounters two significant challenges: the divergence of the partial-wave series (PWS) \cite{Dolan_2008a} and the divergence of the Poisson spot \cite{Zhang_Fan_2021}. Through rigorous calculations, this work resolves these divergences by analyzing \emph{scalar} wave scattering in two distinct gravitational backgrounds: a weak-field body described by Newtonian potentials and a BH with strong gravity governed by the Regge-Wheeler (RW) potential. The computational approach for scalar scattering is also applicable to solving the GW scattering process, which is left as our future work.

Before the formal analysis, we would like to outline the standard partial-wave method for planar scalar wave scattering in spherically symmetric backgrounds. The computational procedure comprises three key steps: (1) One decomposes the planar incident wave in terms of spherical harmonics $Y_{\ell m}(\theta,\varphi)$ in the frequency domain, where $\ell$ (angular quantum number) and $m$ (magnetic quantum number) characterize the multipole moments. The incident wave is then written as a summation over all $(\ell,m)$ modes, which is named as PWS. The expanding coefficients are referred to as the radial function. (2) One asymptotically expands the radial function in the far-field regime and splits it into ingoing (incident) and outgoing (reflected) waves, with corresponding incident and reflection coefficients. The incident coefficient is kept unchanged, and the reflection coefficient for the scattered wave is determined by solving the radial equation, e.g., the Coulomb wave equation for Newtonian or Coulomb potentials or the spin-0 RW equation for the BH background. (3) After obtaining the asymptotic radial function, one finally reconstructs the full scattered wave field through resumming the PWS.

The first computational challenge arises from the divergence of the asymptotic PWS \cite{Dolan_2008a}. Several regularization techniques have been employed to address this divergence, including the series reduction method (SRM) \cite{Yennie_1954,Stratton_2020,Dolan_2008b}, Ces\`{a}ro summation \cite{Pijnenburg_2024b}, the complex angular momentum method \cite{Andersson_1994,folacci_2019a,folacci_2019b}, and the Fresnel half-wave-zone method \cite{Zhang_Fan_2021}. While these methods prove effective in regions distant from the optical axis (the symmetric axis of the scattering process), they universally fail to resolve the divergence near the optical axis, that is, the second divergence we encountered, causing a Poisson spot with infinite luminosity \cite{Zhang_Fan_2021}. Consequently, the absence of a robust regularization scheme for these two divergences hinders the derivation of a physically meaningful and accurate waveform in the near-axis region.

This work systematically investigates these two aforementioned divergences in partial-wave analysis. When placing the observer at a finite radius and avoiding the asymptotic expansion, the convergent PWS and Poisson spot are obtained without any regularization scheme. We present numerical computations for the scattered wave and Poisson spot in Newtonian and BH scattering. The adopted scheme is compared with that by previous studies, e.g., \cite{Chan_2025}, where non-ignorable deviation appears in the near-axis region. This demonstrates the necessity of our rigorous calculations to construct lensing waveforms accurately, especially near the optical axis. Finally, it is shown that our calculation gives consistent results with the diffraction integral \cite{Takahashi_2003}.

The paper is structured as follows: Section \ref{sec:inc} reviews the partial-wave analysis for planar incident waves. Section \ref{sec:Newton} focuses on weak-field scattering in the Newtonian potential and discusses the cause of DCS divergences. Section \ref{sec:RW} extends the above analysis to BH scattering. A summary and further discussion are included in Section \ref{sec:summary}. Throughout the paper, we work in geometric units, $c=G=1$, where $c$ is the speed of light in vacuum and $G$ is the gravitational constant.

\section{\label{sec:inc}Incident scalar plane wave}
As the simplest setup, we set the incident wave as a plane wave, with the time-domain wave function expressed as
\begin{equation}
\label{eq-2:incident-wavefunction}
\psi_{\rm{inc}}(t,\bm{r})
=\frac{1}{2\pi}\int_{-\infty}^{\infty}\mathcal{A}(k)\tilde{\psi}_{\rm{inc}}(k,\bm{r})e^{-ikt}\dd k,
\end{equation}
where the frequency-domain waveform is
\begin{equation}
\label{eq-2:incident-wavefunction-k}
\tilde{\psi}_{\rm{inc}}(k,\bm{r})
=e^{ikr\cos\theta}.
\end{equation}
In the above expressions, $t$ is the global coordinate time, and $\bm{r}\equiv(x,y,z)$ represents the spatial Cartesian coordinates, with corresponding spherical-polar coordinates $(r,\theta,\varphi)$. We set the wave to propagate along the $z$-axis, with $k$ being the wavenumber or angular frequency. For a monochromatic wave with unit amplitude, Eq.\,(\ref{eq-2:incident-wavefunction}) simplifies by replacing the frequency integral with
\begin{equation}
\frac{1}{2\pi}\int_{-\infty}^{\infty}\mathcal{A}(k)\dd k\rightarrow1.
\end{equation}
For simplicity, this work focuses exclusively on frequency-domain analysis.

Due to the completeness of $Y_{\ell m}(\theta,\varphi)$, which represents the eigenstates of orbital angular momentum, an arbitrary scalar field can be decomposed as 
\begin{equation}
\label{eq-2:plane-wave-spherical-harmonics-expansion}
\tilde{\psi}_{\rm{inc}}(k,\bm{r})
=\sum_{\ell=0}^{\infty}\tilde{a}_{\ell}(k,r)\mathrm{P}_{\ell}(\cos\theta),
\end{equation}
where $\mathrm{P}_{\ell}(\cos\theta)$ is the $\ell$-th order Legendre polynomial of the first kind. This is related to spherical harmonics through $Y_{\ell0}(\theta)=((2\ell+1)/4\pi)^{1/2}\mathrm{P}_{\ell}(\cos\theta)$, with the azimuth dependence omitted due to the axial symmetry of the system. The expansion coefficient $\tilde{a}_{\ell}(k,r)$ are given by
\begin{equation}
\label{eq-2:expansion-coefficients}
\tilde{a}_{\ell}(k,r)
=i^{\ell}(2\ell+1)j_{\ell}(kr),
\end{equation}
where $j_{\ell}(kr)$ represents the $\ell$-th order spherical Bessel function of the first kind. 

The spherical-harmonics expansion (\ref{eq-2:plane-wave-spherical-harmonics-expansion}) is valid for any finite $r$. At the large distance, the asymptotic expansion of {\color{red}$\tilde{a}_{\ell m}(k,r)$} is given by
\begin{equation}
\label{eq-2:expansion-coefficients-asymptotic-expansion}
\tilde{a}_{\ell}(k,r\rightarrow\infty)\rightarrow(-1)^{\ell+1}\frac{(2\ell+1)}{2ikr}\Big\{e^{-ikr}-(-1)^{\ell}e^{ikr}\Big\},
\end{equation}
including two linear-independent terms, the ingoing wave $e^{-ikr}$ and outgoing wave $e^{ikr}$. To derive Eq.\,(\ref{eq-2:expansion-coefficients-asymptotic-expansion}), We have used the large-$kr$ expansion of $j_{\ell}(z)$,
\begin{equation}
\label{eq-2:spherical-Bessel-asymptotic-expansion}
j_{\ell}(z)\rightarrow z^{-1}\sin\left(z-\frac{\ell\pi}{2}\right).
\end{equation}
And then Eq.\,(\ref{eq-2:plane-wave-spherical-harmonics-expansion}) is rewritten as
\begin{equation}
\label{eq-2:plane-wave-spherical-harmonics-expansion-asymptotic}
\begin{aligned}
\tilde{\psi}_{\rm{inc}}(k,\bm{r})
&=\sum_{\ell=0}^{\infty}
(-1)^{\ell+1}\frac{(2\ell+1)}{2ikr}\\
&\times\Big\{e^{-ikr}-(-1)^{\ell}e^{ikr}\Big\}
\mathrm{P}_{\ell}(\cos\theta),
\end{aligned}
\end{equation}
for large $kr$. Eq.\,(\ref{eq-2:plane-wave-spherical-harmonics-expansion-asymptotic}) describes the behavior of the scalar waves without scattering, which provides a reasonable boundary condition for solving the scattering wave function in the subsequent sections. This work mainly focuses on the scattering in the isolated Schwarzschild spacetime. Due to its spherical symmetry, the spherical harmonic $Y_{\ell m}(\theta,\varphi)$ [or the Legendre function $\mathrm{P}_{\ell}(\cos\theta)$] remains the eigenstates of the conserved orbital angular momentum. The external gravitational field changes only the radial wave function $\tilde{a}_{\ell}(k,r)$. From the perspective of asymptotic infinity, only the outgoing coefficient [before $e^{ikr}$ of Eq.\,(\ref{eq-2:expansion-coefficients-asymptotic-expansion})] is modified, because of the wave reflection by the BH potential barrier, and the incident coefficient [before $e^{-ikr}$ of Eq.\,(\ref{eq-2:expansion-coefficients-asymptotic-expansion})] remains unchanged.

\begin{figure}[H]
    \centering
    \includegraphics[width=\linewidth]{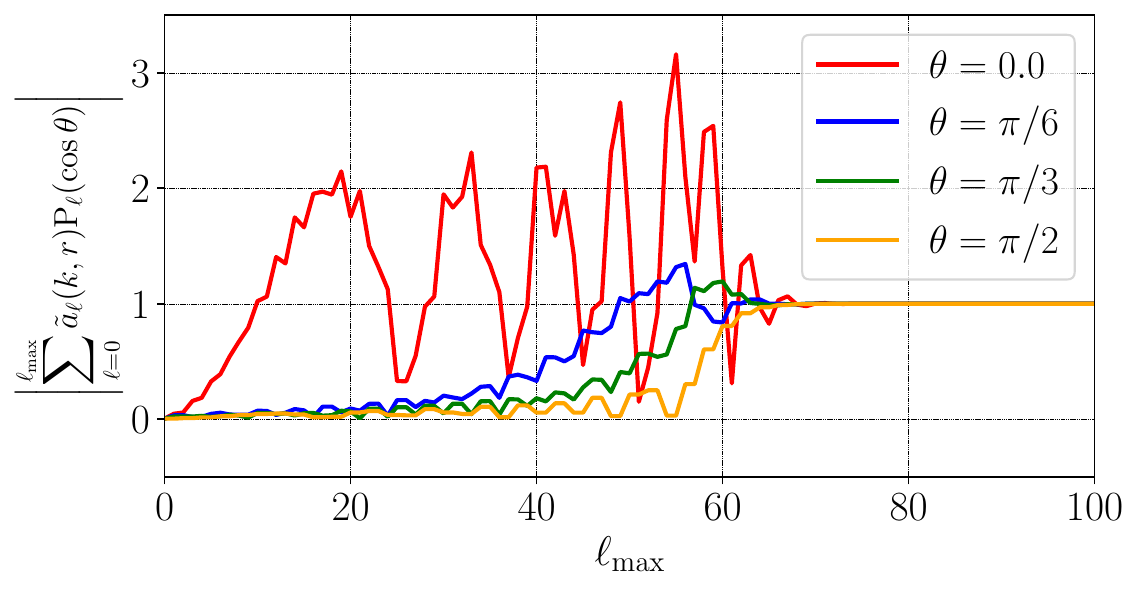}
    \caption{The summation and convergence of Eq.\,(\ref{eq-2:plane-wave-spherical-harmonics-expansion}) for a monochromatic planar scalar wave, with $kr=60.0$ and $\theta=\{0,\pi/6,\pi/3,\pi/2\}$. The truncation is $\ell_{\max}\sim kr$ approximately.}
    \label{fig:plane_wave_convergence}
\end{figure}

\begin{widetext}

\begin{figure}[H]
    \centering
    \includegraphics[width=0.90\textwidth]{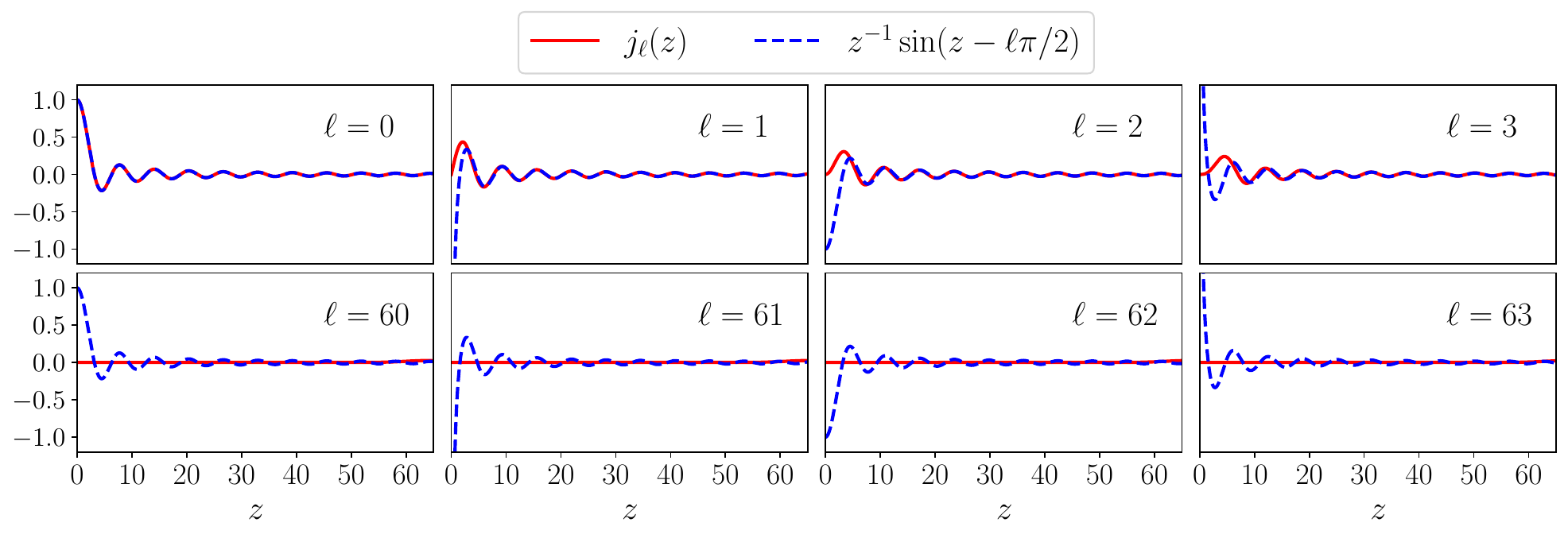}
    \caption{Comparison between the spherical Bessel function $j_{\ell}(z)$ and its large-$kr$ expansion for low and high $\ell$ modes. The region we focus on is around $z\sim60.0$. The low-$\ell$ modes (e.g., $\ell=0,1,2,3\ll 60.0$) are well approximated by their large-$kr$ expansion. But this is invalid for the high-$\ell$ modes (e.g., $\ell=60,61,62,63\sim 60.0$).}
    \label{fig:spherical_Bessel_asymptotic_expansione}
\end{figure}
\end{widetext}

Before moving to the scattering process, it is meaningful to glance at the convergence of the above PWS. The right-hand side of Eq.\,(\ref{eq-2:plane-wave-spherical-harmonics-expansion}) is convergent for an arbitrary finite $r$, and the truncation can be empirically taken as $\ell_{\max}\sim kr$, as shown in Fig.\,\ref{fig:plane_wave_convergence}. However, it is not hard to find that the right-hand side of Eq.\,(\ref{eq-2:plane-wave-spherical-harmonics-expansion-asymptotic}) is divergent. This is because that the asymptotic expansion of $j_{\ell}(z)$ [see Eq.\,(\ref{eq-2:spherical-Bessel-asymptotic-expansion})] is valid only for $kr\gg\ell$ rather than $kr\gg1$ [see Fig.\,\ref{fig:spherical_Bessel_asymptotic_expansione}]. For large but finite $kr$, all of the eigenmodes with $\ell\lesssim\ell_{\max}\sim kr$ contribute to the total wave function. The expansion (\ref{eq-2:plane-wave-spherical-harmonics-expansion-asymptotic}) is a valid approximation only for low-$\ell$ modes (e.g., $\ell\ll\ell_{\max}$), but not for high-$\ell$ modes (e.g., $\ell\lesssim\ell_{\max}$). This is the cause of the divergence in Eq.\,(\ref{eq-2:plane-wave-spherical-harmonics-expansion-asymptotic}).

\section{\label{sec:Newton}Newtonian scattering: Weak-field approximated solution}
\subsection{Wave equation}
This section considers the scattering process by an isolated, weak-gravity, and spherically symmetric scatterer. Such a scatterer possesses rest mass $M$ and is located at the origin. It is noted that an analytical solution exists for this scattering model, which provides a convenient comparison with the partial-wave method. In the Cartesian coordinate $[t,\bm{r}=(x,y,z)]$, its metric is written as
\begin{equation}
\label{eq-3:background}
g_{\mu\nu}\dd x^{\mu}\dd x^{\nu}=-\Big[1+2\phi(r)\Big]\dd t+\Big[1-2\phi(r)\Big]\dd\bm{r}^2,
\end{equation}
in the weak-field approximation, where $\phi(r)=-M/r$ is the Newtonian potential. Our starting point is the massless Klein-Gordon (KG) equation on curved backgrounds. 
\begin{equation}
\label{eq-3:KG}
\Box^2\psi(t,\bm{r})
=\frac{1}{\sqrt{-g}}
\partial_{\mu}\left[\sqrt{-g}
g^{\mu\nu}\partial_{\nu}\right]\psi(t,\bm{r})=0.
\end{equation}
$\Box^2\equiv\nabla_{\alpha}\nabla^{\alpha}$ is the d'Alembert operator compatible with the background, and $g=\det|g_{\mu\nu}|$ is the determinant of the metric. Working with the Cartesian coordinates, we simplify Eq.\,(\ref{eq-3:KG}) as \cite{Peters_1974}
\begin{equation}
\label{eq-3:KG-weak-field}
\bm{\nabla}^2\psi(t,\bm{r})
=(1-4\phi)\frac{\partial^2}{\partial t^2}\psi(t,\bm{r}),
\end{equation}
up to the linear order of $\phi(r)$. $\bm{\nabla}^2=\partial_i\partial_{i}$ is the Laplacian operator.

Through numerical simulation, Ref.\,\cite{He_2021} solved Eq.\,(\ref{eq-3:KG-weak-field}) in the time domain and presented the interference of scattering waves. Eq.\,(\ref{eq-3:KG-weak-field}) is also the theoretical foundation of the diffraction integral, the predominant computational tool for gravitational-lensed GW signals \cite{Takahashi_2003}. In this section, we will present the analytical solution to Eq.\,(\ref{eq-3:KG-weak-field}) in the frequency domain. Two different approaches will be applied, and two equivalent exact solutions will be shown.

\subsection{\label{subsec:3-B}Solution in paraboloidal coordinate}
\begin{widetext}

\begin{figure}[H]
    \centering
    \begin{subfigure}[b]{0.24\textwidth}
        \includegraphics[width=\linewidth]{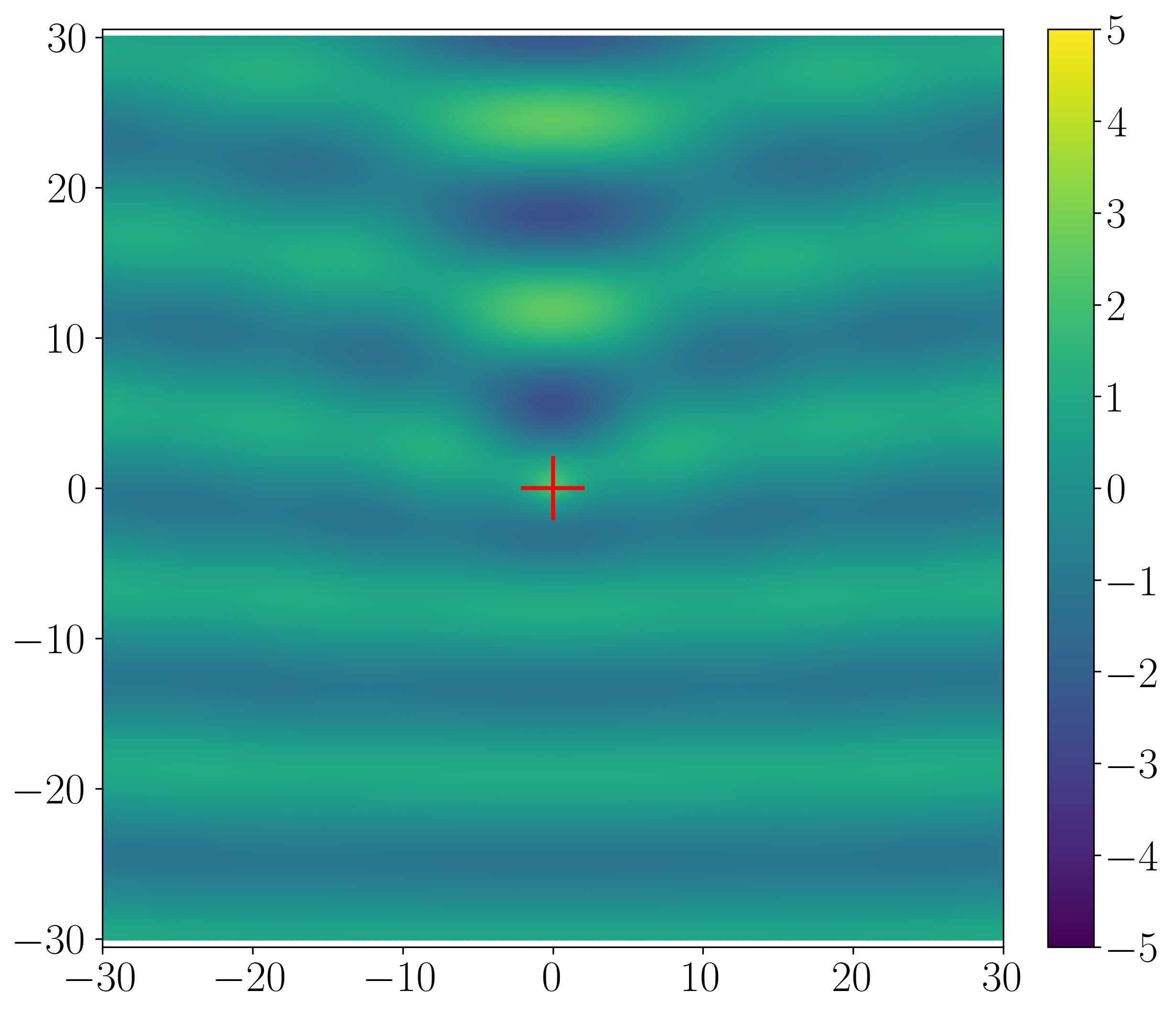}
        \caption{$\mathrm{Re}\,\Big\{\tilde{\psi}(k=0.5/M,\bm{r})\Big\}$}
    \end{subfigure}
    \begin{subfigure}[b]{0.24\textwidth}
        \includegraphics[width=\linewidth]{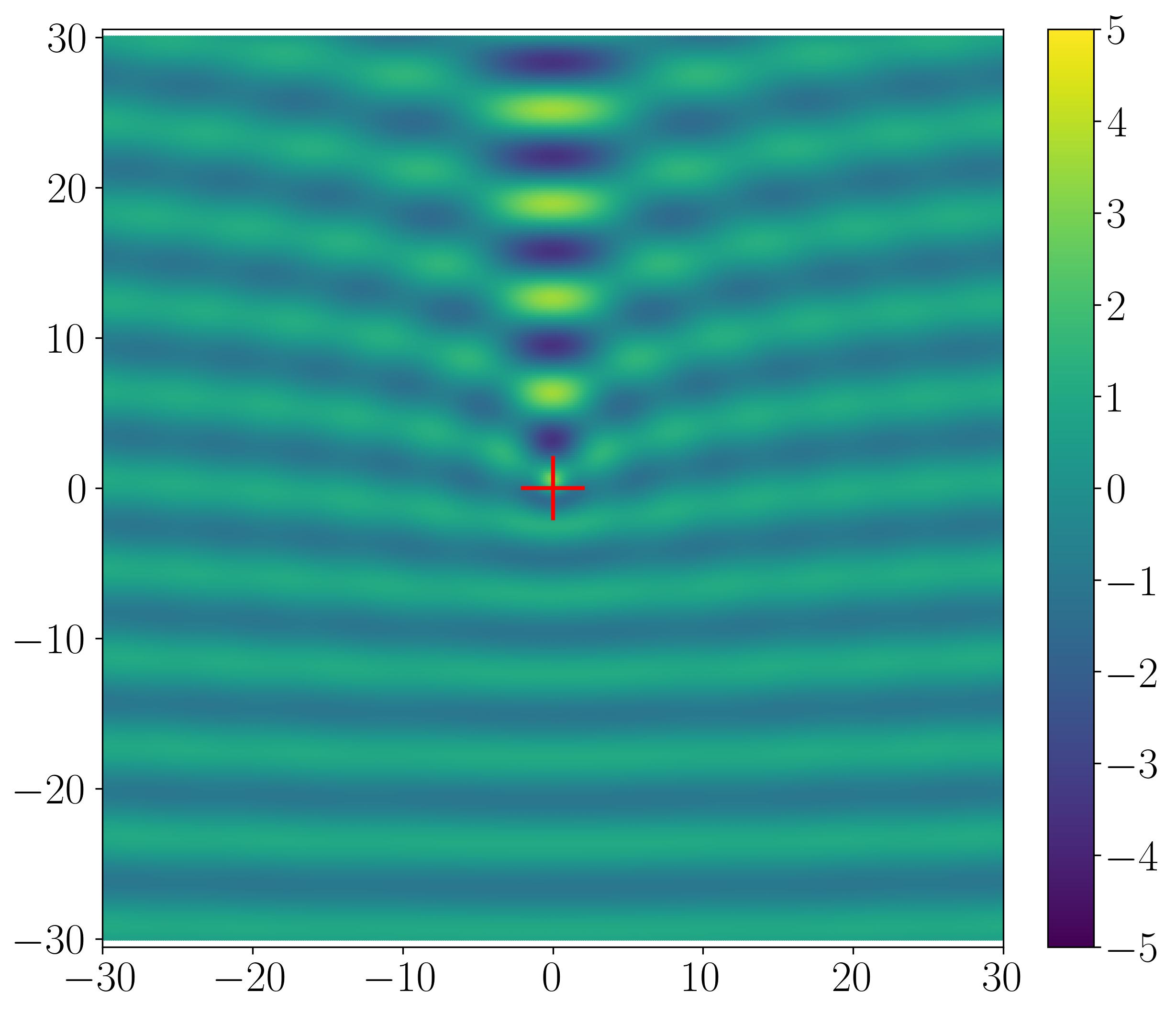}
        \caption{$\mathrm{Re}\,\Big\{\tilde{\psi}(k=1.0/M,\bm{r})\Big\}$}
    \end{subfigure}
    \begin{subfigure}[b]{0.24\textwidth}
        \includegraphics[width=\linewidth]{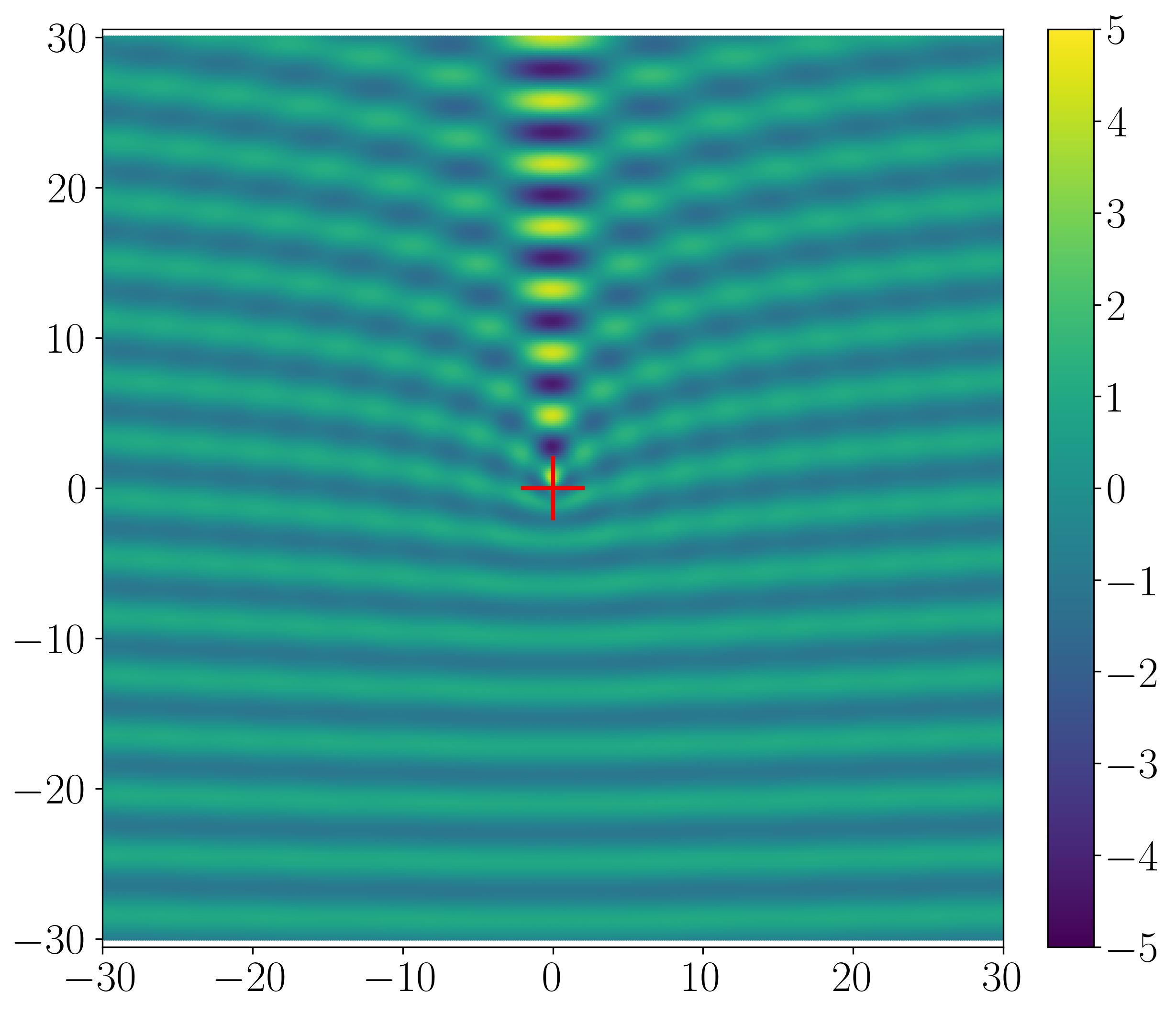}
        \caption{$\mathrm{Re}\,\Big\{\tilde{\psi}(k=1.5/M,\bm{r})\Big\}$}
    \end{subfigure}
    \begin{subfigure}[b]{0.24\textwidth}
        \includegraphics[width=\linewidth]{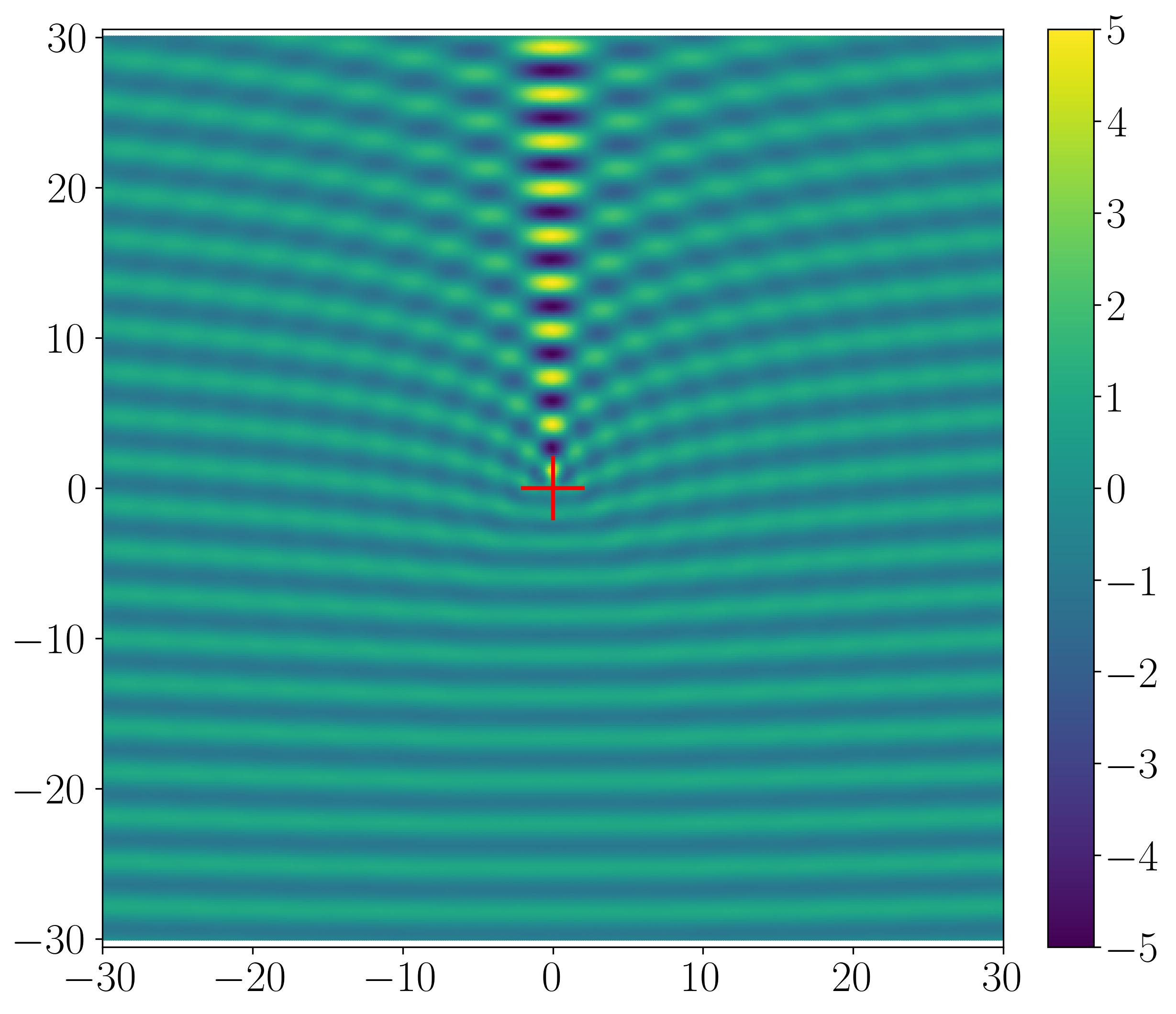}
        \caption{$\mathrm{Re}\,\Big\{\tilde{\psi}(k=2.0/M,\bm{r})\Big\}$}
    \end{subfigure}

    \begin{subfigure}[b]{0.24\textwidth}
        \includegraphics[width=\linewidth]{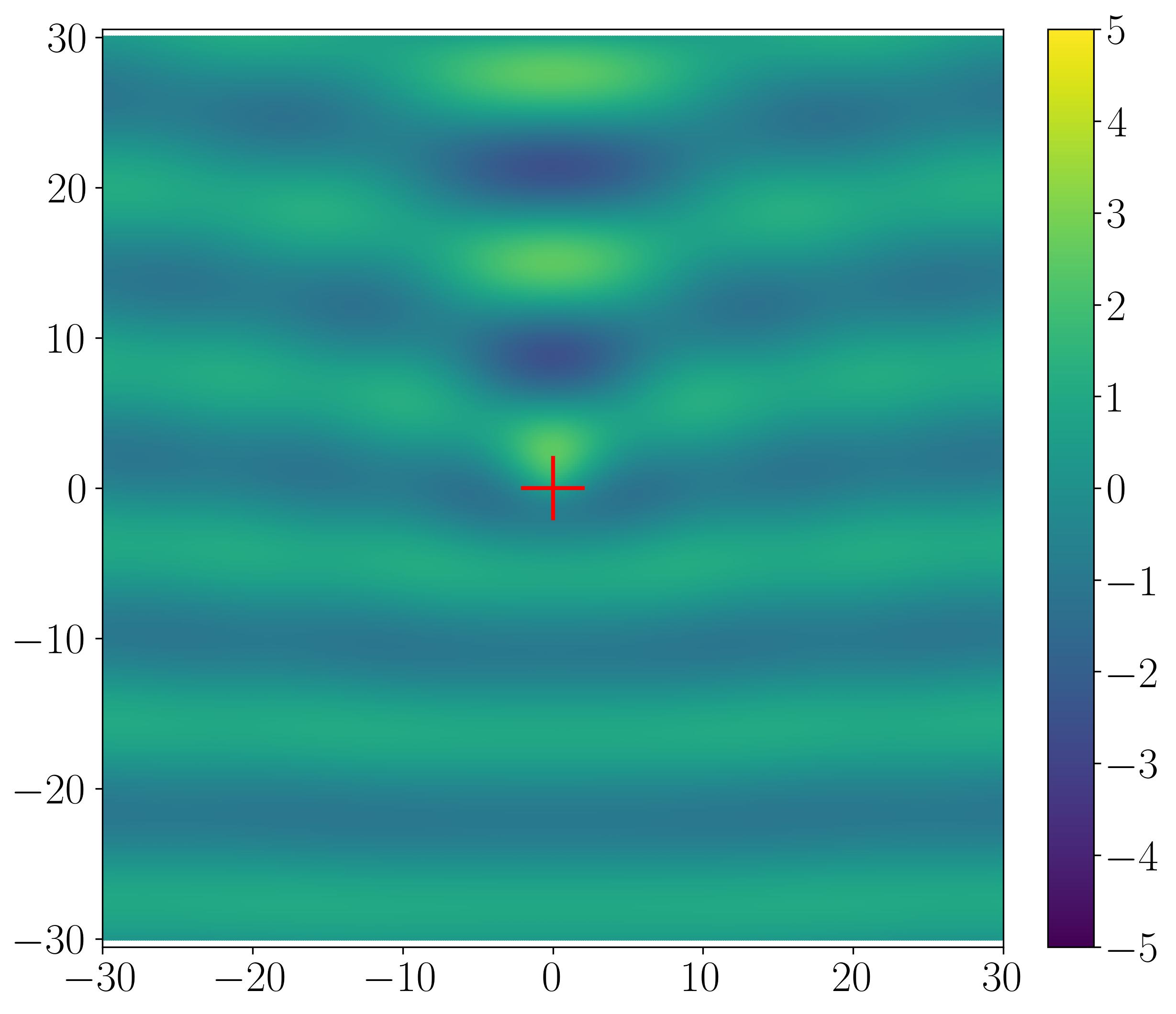}
        \caption{$\mathrm{Im}\,\Big\{\tilde{\psi}(k=0.5/M,\bm{r})\Big\}$}
    \end{subfigure}
    \begin{subfigure}[b]{0.24\textwidth}
        \includegraphics[width=\linewidth]{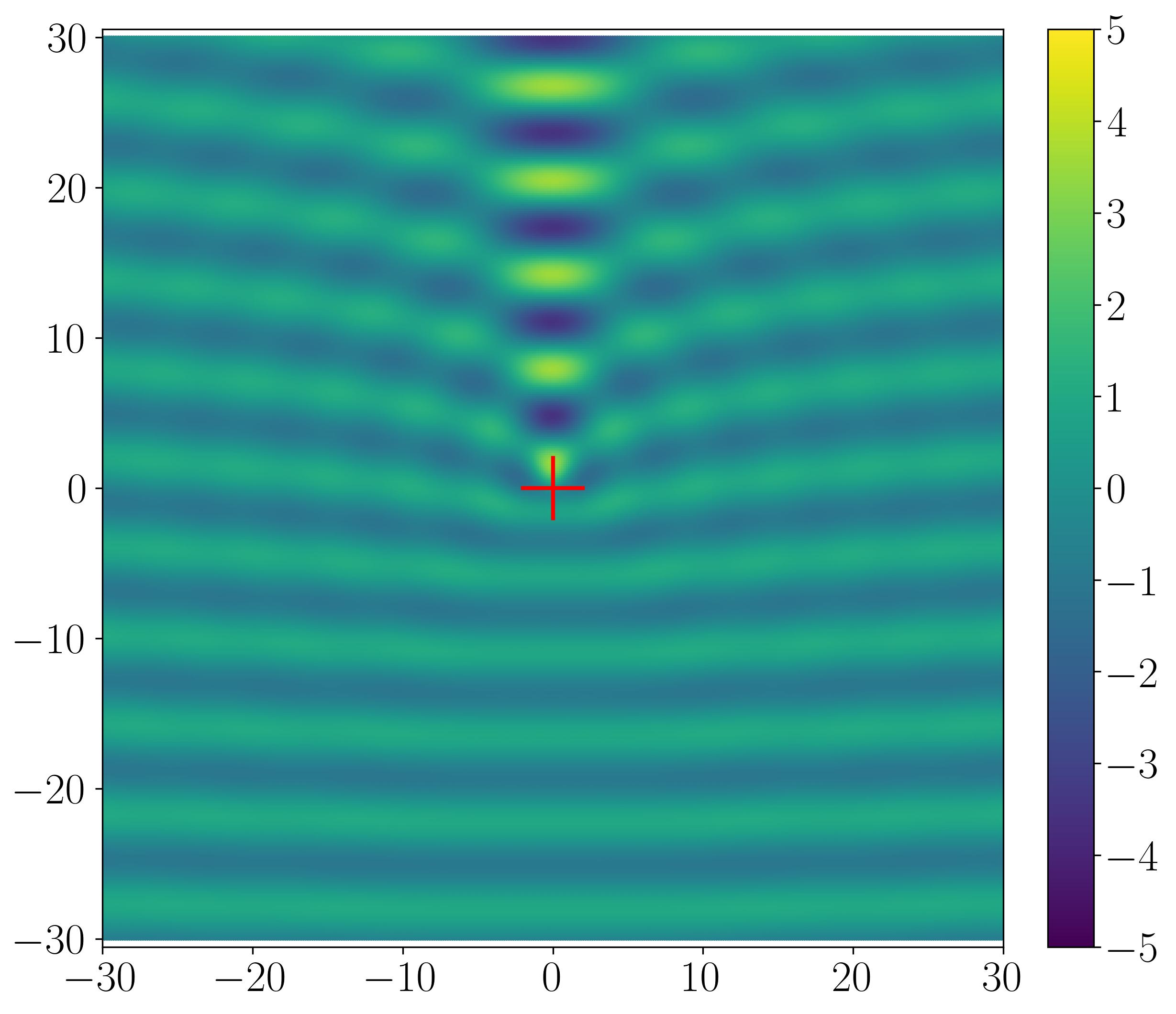}
        \caption{$\mathrm{Im}\,\Big\{\tilde{\psi}(k=1.0/M,\bm{r})\Big\}$}
    \end{subfigure}
    \begin{subfigure}[b]{0.24\textwidth}
        \includegraphics[width=\linewidth]{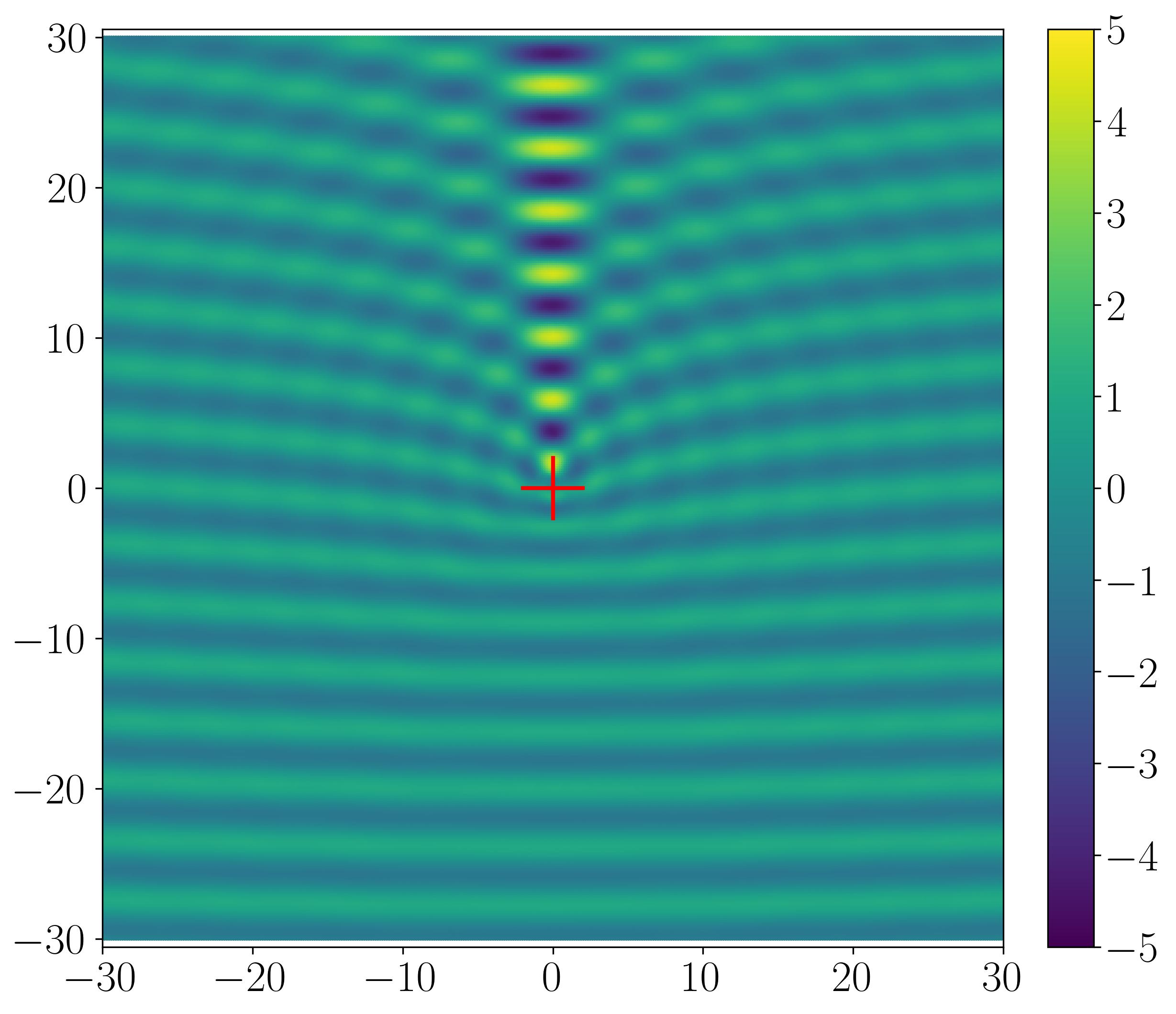}
        \caption{$\mathrm{Im}\,\Big\{\tilde{\psi}(k=1.5/M,\bm{r})\Big\}$}
    \end{subfigure}
    \begin{subfigure}[b]{0.24\textwidth}
        \includegraphics[width=\linewidth]{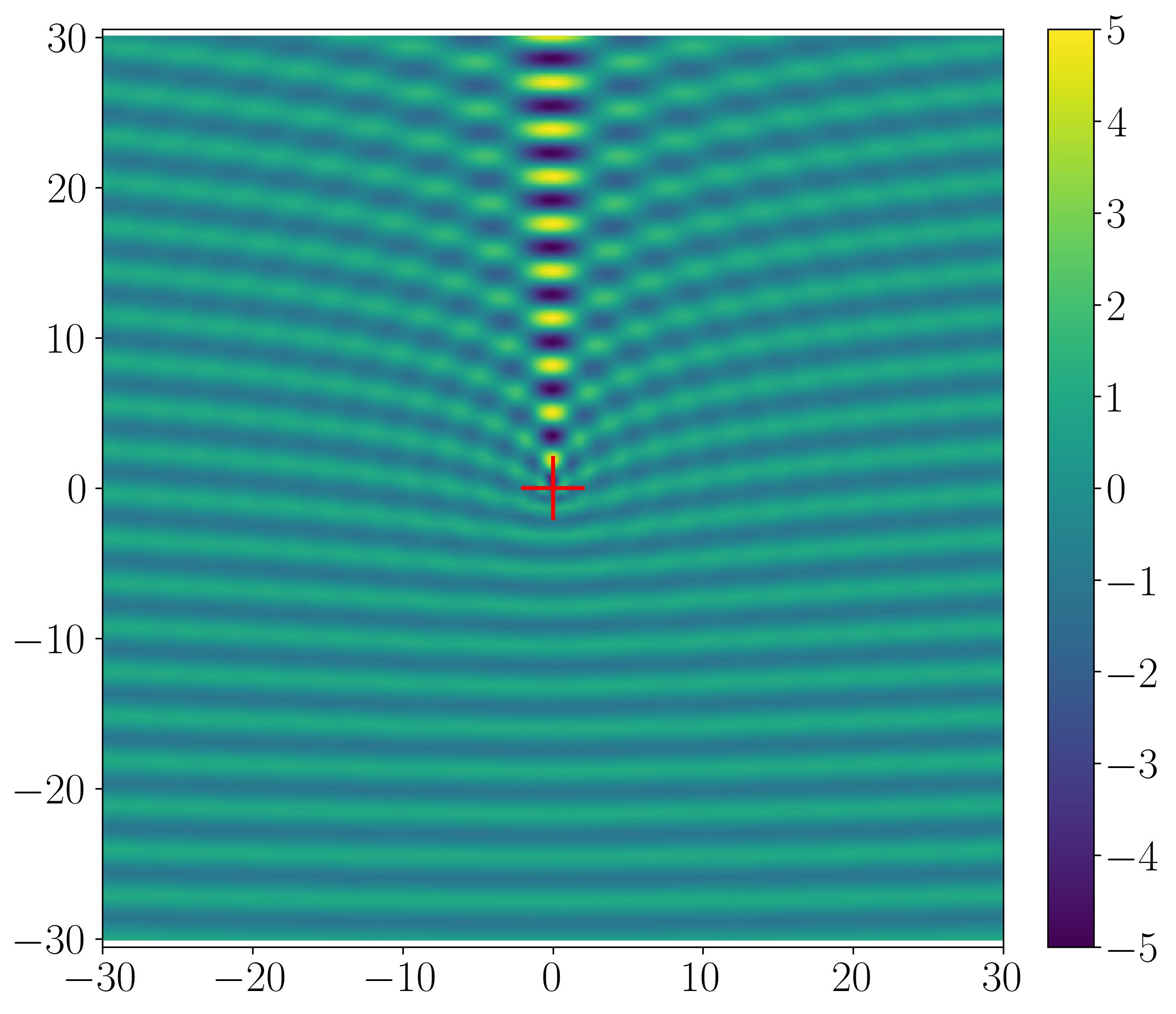}
        \caption{$\mathrm{Im}\,\Big\{\tilde{\psi}(k=2.0/M,\bm{r})\Big\}$}
    \end{subfigure}
    \caption{Scattered monochromatic scalar wave fields calculated through Eq.\,(\ref{eq-3:scattering-wave-function-paraboloidal}) with $k=\{0.5,1.0,1.5,2.0\}/M$. The abscissa ($x$-axis) and ordinate ($z$-axis) in these figures are in units of BH mass. The red crosses in the figures represent the position of the scatterer.}
    \label{fig:scattering-wave-Newton}
\end{figure}

\end{widetext}

The first approach to solving the Eq.\,(\ref{eq-3:KG-weak-field}) works with the paraboloidal coordinates $\{\xi,\eta,\varphi\}$, relating with the Cartesian coordinates by
\begin{equation}
x=\sqrt{\xi\eta}\cos\varphi,\,
y=\sqrt{\xi\eta}\sin\varphi,\,
z=\frac{1}{2}(\xi-\eta).
\end{equation}
The major advantage of this method is that it does not rely on spherical-harmonic decomposition, providing a way to compare with the partial-wave method. Taking the Fourier transform
\begin{equation}
\label{eq-3-wave-function-Fourier}
\psi(t,\bm{r})=\frac{1}{2\pi}\int_{-\infty}^{\infty}\mathcal{A}(k)\dd ke^{-ikt}\tilde{\psi}(k,\bm{r})
\end{equation}
and transform $\bm{\nabla}^2$ into paraboloidal coordinates, we re-express the wave equation as \cite{He_2021,Pijnenburg_2024a}
\begin{equation}
\label{eq-3:wave-equation-paraboloidal}
\begin{aligned}
&\Bigg\{\frac{4}{\xi+\eta}\left[\frac{\partial}{\partial \xi}\left(\xi \frac{\partial}{\partial \xi}\right)+\frac{\partial}{\partial \eta}\left(\eta \frac{\partial}{\partial \eta}\right)\right]\\
&\qquad+\frac{1}{\xi \eta} \frac{\partial^2}{\partial \varphi^2}+k^2\left(1+\frac{8M}{\xi+\eta}\right)\Bigg\}\tilde{\psi}(k,\bm{r})=0.
\end{aligned}
\end{equation}
In the flat-background limit, where $M=0$, the solution to $\tilde{\psi}(k,\bm{r})$ reduces to the plane wave, $\tilde{\psi}_{\rm inc}(k,\bm{r})$ (\ref{eq-2:incident-wavefunction-k}). Since the whole scattering process remains axisymmetric about the optical axis, we require the solution to be independent of the azimuth angle $\varphi$, and assume the variable-separated solution to Eq.\,(\ref{eq-3:wave-equation-paraboloidal}) to be \cite{He_2021,Pijnenburg_2024a}
\begin{equation}
\label{eq-3:wave-function-paraboloidal-variable-separated}
\tilde{\psi}(k,\bm{r})\propto e^{ik(\xi-\eta)/2}W(\eta).
\end{equation}
remaining as an undetermined coordinate-independent overall constant. The function $W(\eta)$ satisfies
\begin{equation}
\label{eq-3:hypergeometric-function}
\eta W''(\eta)+(1-ik\eta)W'(\eta)-k\gamma W(\eta)=0,
\end{equation}
where $\gamma\equiv-2Mk$, and we can express its solution in terms of the Kummer hypergeometric function \footnote{Two linearly independent solutions to confluent hypergeometric equation (\ref{eq-3:hypergeometric-function}) are Kummer function ${_1F_1}(-i\gamma,1;ik\eta)$ and Tricomi function $U(-i\gamma,1;ik\eta)$. We require the wave function to have a finite value at the origin, and therefore only the Kummer function should be considered because $U(b,c,z)$ is singular at $z=0$.}
\begin{equation}
W(\eta)={_1F_1}\Big\{-i\gamma,1;ik\eta\Big\}.
\end{equation}
To determine the overall constant, we consider the asymptotic behavior of Eq.\,(\ref{eq-3:wave-function-paraboloidal-variable-separated}) along the backward direction, where
$\theta=\pi$ and $\eta=2r$. For large $kr$, Eq.\,(\ref{eq-3:wave-function-paraboloidal-variable-separated}) is 
\begin{equation}
\label{eq-3:wave-function-paraboloidal-variable-separated-asymptotic}
\tilde{\psi}(k,\bm{r})\propto\frac{e^{\pi\gamma/2}}{\Gamma(1+i\gamma)} \left\{e^{-ikr_*} +\frac{1}{2ikr}\frac{\Gamma(1+i\gamma)}{\Gamma(-i\gamma)}e^{ikr_*}\right\}.
\end{equation}
The ingoing/outgoing wave is represented by $e^{\pm ikr_*}$ rather than $e^{\pm ikr}$, where 
\begin{equation}
\label{eq-3-r-star}
r_* \equiv r-\frac{\gamma}{k}\ln(2kr)
\end{equation}
is the tortoise coordinate, reflecting the long-range nature of the gravitational interaction. In other words, the plane wave is \emph{not} an exact solution to Eq.\,(\ref{eq-3:KG}), but an extra phase shift should be introduced to describe the slight distortion of the wavefront. On the backward direction, the incident plane wave is $\tilde{\psi}_{\rm inc}(k,\bm{r})\rightarrow e^{-ikr}$. Matching the incident coefficient in Eq.\,(\ref{eq-3:wave-function-paraboloidal-variable-separated-asymptotic}) gives the result of overall constant,
\begin{equation}
e^{-\pi\gamma/2}\Gamma(1+i\gamma).
\end{equation}
In summary, the scattering scalar wave field is exactly described by the wave function
\begin{equation}
\label{eq-3:scattering-wave-function-paraboloidal}
\begin{aligned}
\tilde{\psi}&(k,\bm{r})=e^{ikr\cos\theta}e^{-\pi\gamma/2}\\
&\times\Gamma(1+i\gamma){_1F_1}\Big\{-i\gamma,1;2ikr\sin^2(\theta/2)\Big\}.
\end{aligned}
\end{equation}
Fig.\,\ref{fig:scattering-wave-Newton} shows the numerical result of Eq.\,(\ref{eq-3:scattering-wave-function-paraboloidal}) for some specific values of $k$. 

In the negative-$z$ region, the wave field behaves roughly as a plane wave. However, in the positive-$z$ region, its behavior is rather complicated. In the region far from the optical axis, the wave field can be seen as the linear superposition of a distorted plane wave and an outgoing spherical wave centered on the scatterer. In the region close to the optical axis, there is a strong interference effect, and a bright Poisson spot forms. From Eq.\,(\ref{eq-3:scattering-wave-function-paraboloidal}), one finds that the brightness of the Poisson spot is a constant
\begin{equation}
\lim_{\theta\rightarrow0^+}\tilde{\psi}(k,\bm{r})=e^{-\pi\gamma/2}\Gamma(1+i\gamma),
\end{equation}
along the positive half $z$ axis, which is independent of distance, but depends on the frequency, indicating that a distant observer still observes this bright Poisson spot.

This is because of the assumption of a plane wave, which consists of infinite partial waves, ranging from $\ell=0$ to $\ell\rightarrow\infty$. For the observer close to the scatterer, the scattering waves are contributed by the low-$\ell$ modes. For the distant observer, the scattered waves still exist due to the contribution of high-$\ell$ modes. On the contrary, if one considers a wave source located at a finite distance, the incident waves comprise only a finite number of eigenmodes of $\ell$, e.g., $0\leqslant\ell\lesssim\ell_{\max}$. The empirical truncation is $\sim kr_s$, with $r_s$ being the distance between the scatterer and source \cite{Kubota_2024}. Once the observer-scatterer distance, denoted by $r$, is large enough, e.g., $r\gg r_s$, the interference pattern and Poisson spot disappear because of the absence of high-$\ell$ modes.

\subsection{\label{subsec:3-C}Partial-wave method}
Another independent approach to solving Eq.\,(\ref{eq-3:KG-weak-field}) is based on the spherical-harmonics decomposition. Similar to what we did for the incident plane wave, we consider the PWS for the frequency-domain wave function as follows,
\begin{equation}
\label{eq-3:scattering-wave-function-spherical-harmonics-expansion}
\tilde{\psi}(k,\bm{r})=
\sum_{\ell=0}^{\infty}\tilde{R}_{\ell}(k,r)\mathrm{P}_{\ell}(\cos\theta).
\end{equation}
The radial function $\tilde{R}_{\ell}(k,r)$ satisfies Coulomb wave equation \cite{NIST:DLMF},
\begin{equation}
\label{eq-3:radial-equation-R-PW}
\tilde{R}_{\ell}''+\frac{2}{r}\tilde{R}_{\ell}'
+\left[k^2-\frac{2\gamma k}{r}-\frac{\ell(\ell+1)}{r^2}\right]\tilde{R}_{\ell}=0,
\end{equation}
where the prime represents the derivative to the radial coordinate $r$. The solution to Eq.\,(\ref{eq-3:radial-equation-R-PW}) is also expressed in terms of the Kummer hypergeometric function, and we arrive at
\begin{equation}
\label{eq-3:radial-solution-R-PW}
\tilde{R}_{\ell}\propto r^{\ell}e^{ikr}{_{1}F_{1}}\Big\{\ell+1+i\gamma,2(\ell+1);-2ikr\Big\},
\end{equation}
with an undetermined overall constant remained.

To determine the overall constant, we asymptotically expand ${_{1}F_{1}}$ at large-$kr$ region, and obtain
\begin{equation}
\label{eq-3:radial-wave-function-c}
\begin{aligned}
&\tilde{R}_{\ell}\propto (-1)^{\ell+1}\frac{\Gamma(2\ell+2)}{\Gamma(\ell+1+i\gamma)}\frac{e^{\pi\gamma/2}}{(2ik)^{\ell}}\\
&\times\frac{1}{2ikr}\left\{e^{-ikr_*}+(-1)^{\ell+1}\frac{\Gamma(\ell+1+i\gamma)}{\Gamma(\ell+1-i\gamma)}e^{ikr_*}\right\}.
\end{aligned}
\end{equation}
which consists of the ingoing and outgoing waves. The harmonics decomposition of the plane wave [see Eq.\,(\ref{eq-2:plane-wave-spherical-harmonics-expansion})] provides boundary conditions for the scattering radial wave function. Except for the extra phase shift $kr\rightarrow kr_*$, the incident coefficient is required to be consistent with that in Eq.\,(\ref{eq-2:expansion-coefficients-asymptotic-expansion}), thus giving
\begin{equation}
\label{eq-3:c-constant}
(2ik)^{\ell}e^{-\pi\gamma/2}\frac{\Gamma(\ell+1+i\gamma)}{\Gamma(2\ell+1)}.
\end{equation}

\begin{figure}[H]
    \centering
    \begin{subfigure}[b]{0.45\textwidth}
        \includegraphics[width=\linewidth]{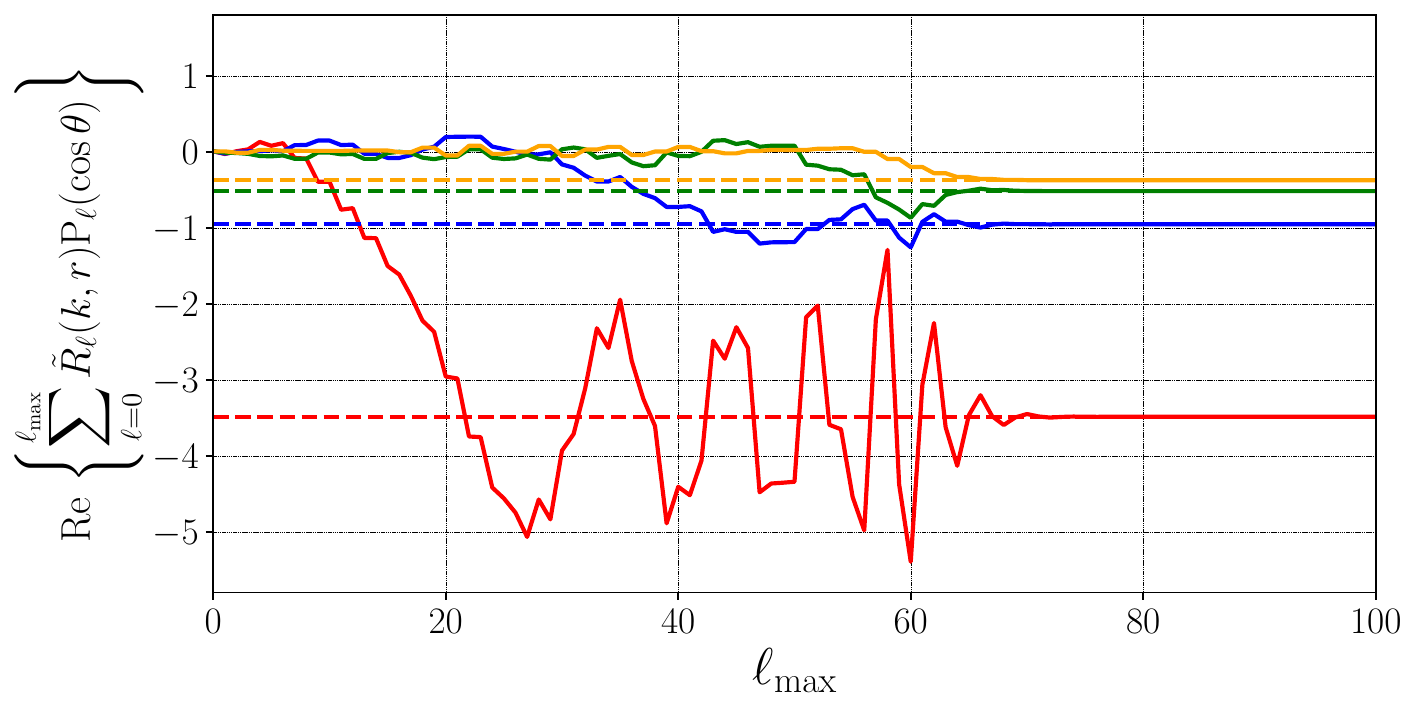}
    \end{subfigure}
    \begin{subfigure}[b]{0.45\textwidth}
        \includegraphics[width=\linewidth]{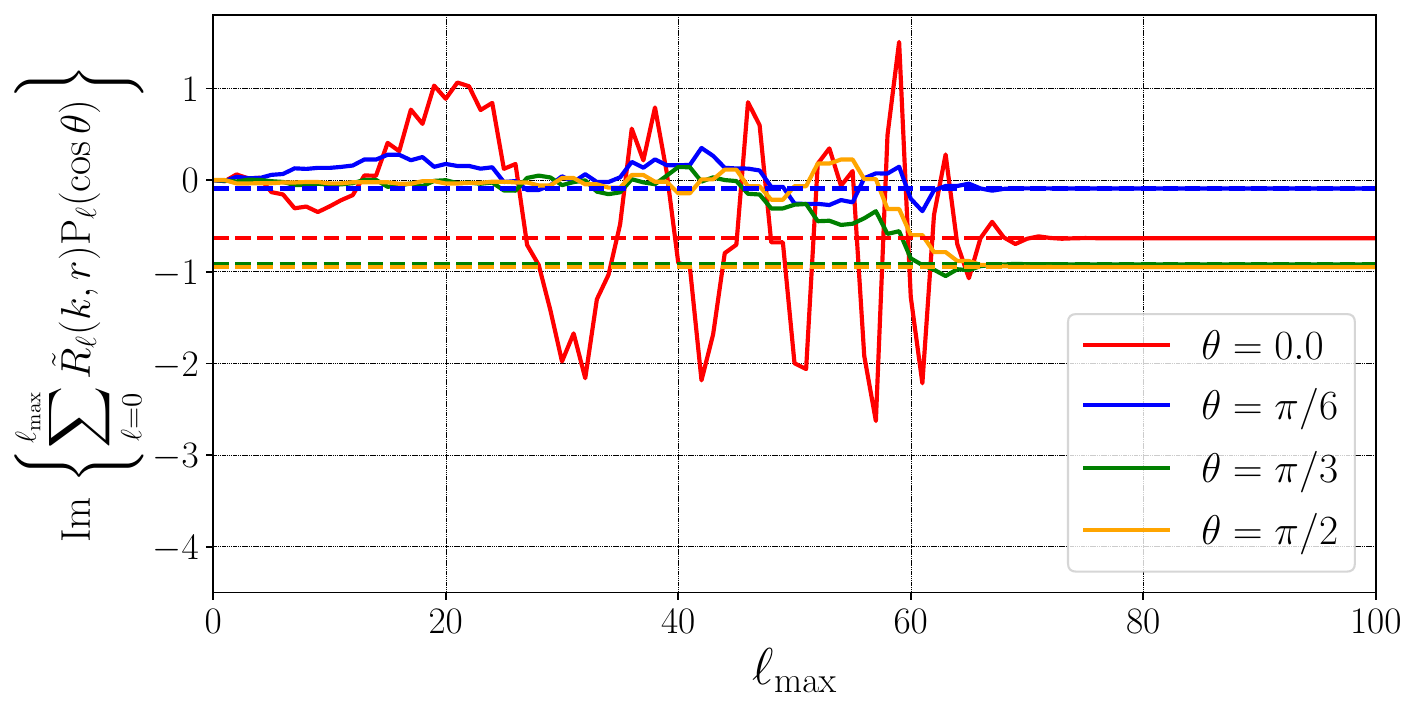}
    \end{subfigure}
    \caption{The convergence of Eq.\,(\ref{eq-3:scattering-wave-function-PW}) and its consistency with paraboloidal-coordinates solution (\ref{eq-3:scattering-wave-function-paraboloidal}). These calculations are implemented for $k=1.0/M$ and at radius $r=60.0M$, therefore the PWS truncation is approximately $\ell_{\max}\sim kr\sim60$. The horizontal dashed line means the values evaluated from Eq.\,(\ref{eq-3:scattering-wave-function-paraboloidal}).}
    \label{fig:Newton-consistency}
\end{figure}

Combining Eqs.\,(\ref{eq-3:scattering-wave-function-spherical-harmonics-expansion},\,\ref{eq-3:radial-solution-R-PW},\,\ref{eq-3:c-constant}), we obtain the normalized radial function as
\begin{equation}
\label{eq-3:scattering-wave-function-PW}
\begin{aligned}
\tilde{R}_{\ell}(k,r)
&=e^{-\pi\gamma/2}
\frac{\Gamma(\ell+1+i\gamma)}{\Gamma(2\ell+1)}(2ikr)^{\ell}e^{ikr}\\
&\times{_{1}F_{1}}\Big\{\ell+1+i\gamma,2(\ell+1);-2ikr\Big\}.
\end{aligned}
\end{equation}
This completed the mathematical derivation of the scalar waves scattered by the Newtonian potential well/barrier. Fig.\,\ref{fig:Newton-consistency} shows the convergence of Eq.\,(\ref{eq-3:scattering-wave-function-PW}) and the consistency between these two different approaches, given by Eqs.\,(\ref{eq-3:scattering-wave-function-paraboloidal}) and (\ref{eq-3:scattering-wave-function-PW}), respectively. 

\subsection{\label{subsec:3-D}Differential cross section and its divergence}
In the last two subsections, we have reviewed the exact solution to Eq.\,(\ref{eq-3:KG-weak-field}). It is noted that Eqs.\,(\ref{eq-3:scattering-wave-function-paraboloidal}) and (\ref{eq-3:scattering-wave-function-spherical-harmonics-expansion}) are always finite for arbitrary space-time points. In quantum theory, a critical observable quantity is DCS, which describes the probability of particles being emitted in a direction away from the scatterer \cite{Sakurai_Napolitano_2020}. The gravitational analogs are usually presented in previous studies \cite{Matzner_1968,Andersson_1995,Glampedakis_2001,Dolan_2008a,Dolan_2008b}.

In the frequency domain, the large-$kr$ approximation of the scattering wave is
\begin{equation}
\label{eq-3:wave-function-variable-separated-asymptotic}
\begin{aligned}
&\tilde{\psi}(k,\bm{r})\rightarrow e^{ikr\cos\theta}e^{i\gamma\ln[2kr\sin^2(\theta/2)]}\\
&+\frac{1}{2ikr\sin^2(\theta/2)}\frac{\Gamma(1+i\gamma)}{\Gamma(-i\gamma)}e^{ikr-i\gamma\ln[2kr\sin^2(\theta/2)]},
\end{aligned}
\end{equation}
for arbitrary $\theta$ with $\sin^2(\theta/2)\gg1/2kr$. The first term represents the incident wave, with long-range phase modulation, and the second term the scattering spherical wave, which is usually written as
\begin{equation}
f(\theta)\frac{e^{ikr_*}}{r},
\end{equation}
where the angular factor is
\begin{equation}
\label{eq-3:angular-factor}
f(\theta)=\frac{1}{2ik}\frac{\Gamma(1+i\gamma)}{\Gamma(-i\gamma)}\Big\{\sin^2(\theta/2)\Big\}^{-2-i\gamma}.
\end{equation}
The DCS is given by \cite{Matzner_1968}
\begin{equation}
\label{eq-3:Rutherford-formula}
\frac{\dd\sigma}{\dd\Omega}=|f(\theta)|^2=\frac{M^2}{\sin^4(\theta/2)}.
\end{equation}
This is the well-known gravitational Rutherford formula, an exact corollary of Eq.\,(\ref{eq-3:scattering-wave-function-paraboloidal}), which possesses a singularity at the forward direction. This is a natural result, because the large-$kr$ expansion of the Kummer hypergeometric function we used in deriving Eq.\,(\ref{eq-3:wave-function-variable-separated-asymptotic}) is only valid for $2kr\sin^2(\theta/2)\gg1$. \emph{In conclusion, the use of asymptotic expansion in the near-axis region is the cause of the forward-scattering singularity}. Except for the near-axis region, Eq.\,(\ref{eq-3:wave-function-variable-separated-asymptotic}) is still an appropriate approximation of Eq.\,(\ref{eq-3:scattering-wave-function-paraboloidal}).

As for the strong-gravity scatterer, e.g., a BH, analytical solutions like Eq.\,(\ref{eq-3:scattering-wave-function-paraboloidal}) are likely to be non-existent. At this point, the partial-wave method becomes the main tool for studying scattering processes. Therefore, studying the DCS in the partial-wave method is beneficial for understanding the BH scattering. The large-$kr$ expansion of Eq.\,(\ref{eq-3:scattering-wave-function-spherical-harmonics-expansion}) is
\begin{equation}
\label{eq-3:wave-function-PW-asymptotic-expansion}
\tilde{\psi}(k,\bm{r})\rightarrow
\Big\{\text{distorted plane wave}\Big\}
+f(\theta)\frac{e^{ikr_*}}{r},
\end{equation}
The distorted plane wave is written as
\begin{equation}
\label{eq-3:wave-function-PW-asymptotic-expansion-plane-wave}
\sum_{\ell=0}^{\infty}
(-1)^{\ell+1}\frac{2\ell+1}{2ikr}
\Big\{e^{-ikr_*}-(-1)^{\ell}e^{ikr_*}\Big\}
\mathrm{P}_{\ell}(\cos\theta).
\end{equation}
The second term of Eq.\,(\ref{eq-3:wave-function-PW-asymptotic-expansion}) represents the scattered spherical wave, with the angular factor being 
\begin{equation}
\label{eq-3:wave-function-PW-asymptotic-expansion-scattered-wave}
f(\theta)=\frac{1}{2ik}
\sum_{\ell=0}^{\infty}
(2\ell+1)
\Big\{e^{2i\delta_{\ell}}-1\Big\}
\mathrm{P}_{\ell}(\cos\theta).
\end{equation}
It should be expected that Eq.\,(\ref{eq-3:wave-function-PW-asymptotic-expansion-scattered-wave}) gives an identical result with Eq.\,(\ref{eq-3:angular-factor}). However, unfortunately, the PWS for the distorted plane wave (\ref{eq-3:wave-function-PW-asymptotic-expansion-plane-wave}) and $f(\theta)$ are both divergent, that is, the ``PWS divergence" mentioned in Section \ref{sec:intro}. This phenomenon has been widely investigated by several works, e.g., Ref.\,\cite{Pijnenburg_2024a}. We have illustrated in Section \ref{sec:inc} that the divergence of Eq.\,(\ref{eq-3:wave-function-PW-asymptotic-expansion-plane-wave}) results from the asymptotic expansion applied for $j_{\ell}(kr)$ with $\ell\sim kr$. The cause for $f(\theta)$ (\ref{eq-3:wave-function-PW-asymptotic-expansion-scattered-wave}) is quiet similar. This is because \emph{the inappropriate asymptotic expansion applied to the radial function}, that is, the large-$kr$ expansion of ${_1F_1}(\cdots,\cdots;\cdots)$ we used in deriving Eq.\,(\ref{eq-3:radial-wave-function-c}) is only valid for $kr\gg\ell$ rather than $kr\gg1$. This observation is shown in Fig.\,\ref{fig:Kummer_asymptotic_expansione}.

To illustrate the physical meaning of this phenomenon, we define \cite{Pijnenburg_2024a}
\begin{equation}
\label{eq-3:R-u}
u_{\ell}(k,r)\equiv r\tilde{R}_{\ell}(k,r),
\end{equation}
and transform the radial equation (\ref{eq-3:radial-equation-R-PW}) into the following Schr\"{o}dinger-like form,
\begin{equation}
\left[\frac{\dd^2}{\dd r^2}+k^2-V^{\rm(N)}_{\ell}(r)\right]u_{\ell}(k,r)=0,
\end{equation}
with the potential being
\begin{equation}
\label{eq-3-Newton-potential}
V^{\rm(N)}_{\ell}(r)=\frac{2\gamma k}{r}-\frac{\ell(\ell+1)}{r^2},
\end{equation}
where the squared frequency $k^2$ plays the role of energy in quantum mechanics. $V^{\rm(N)}_{\ell}(r)$ greater than or less than $k^2$ determines the classical allowed and forbidden regions. The location where $k^2=V^{\rm(N)}_{\ell}(r)$ is referred to as the turning point, denoted by $\hat{r}_{\ell}$ for each $\ell$-order partial wave. Let us suppose a detector to be located at radius $r$, and the detected scattered wave is the superposition of all $\ell$-order partial waves. For low-$\ell$ mode, i.e., $\ell\ll kr$, the corresponding potential function decays rapidly below $k^2$ at $\hat{r}_{\ell}\ll r$, which is shown in Fig.\,(\ref{fig:Newton-potental-low-modes}). Therefore, when arriving at the detector, where $r\gg \hat{r}_{\ell}$, the scattered wave is far away from the barrier/well and is safely regarded as free propagating waves. However, as shown in Fig.\,(\ref{fig:Newton-potental-high-modes}), when $\ell\sim kr$, due to the large centrifugal potential, $k^2$ is approximately equal to the potential, and the detector is close to the turning point. The scattered wave still interacts with the strong potential barrier. Therefore, the above approximation for the low-$\ell$ modes cannot be adopted correctly. This is the reason why the large-$kr$ expansion of the radial function is invalid. Finally, the partial waves with $\ell\gg kr$ are almost fully reflected by the strong potential barrier in the vicinity of the turning point, with $\hat{r}_{\ell}\gg r$, that will not ultimately reach the detector. This is the cause of the natural truncation $\ell_{\max}\sim kr$.

\begin{widetext}

\begin{figure}[H]
    \centering
    \includegraphics[width=0.90\textwidth]{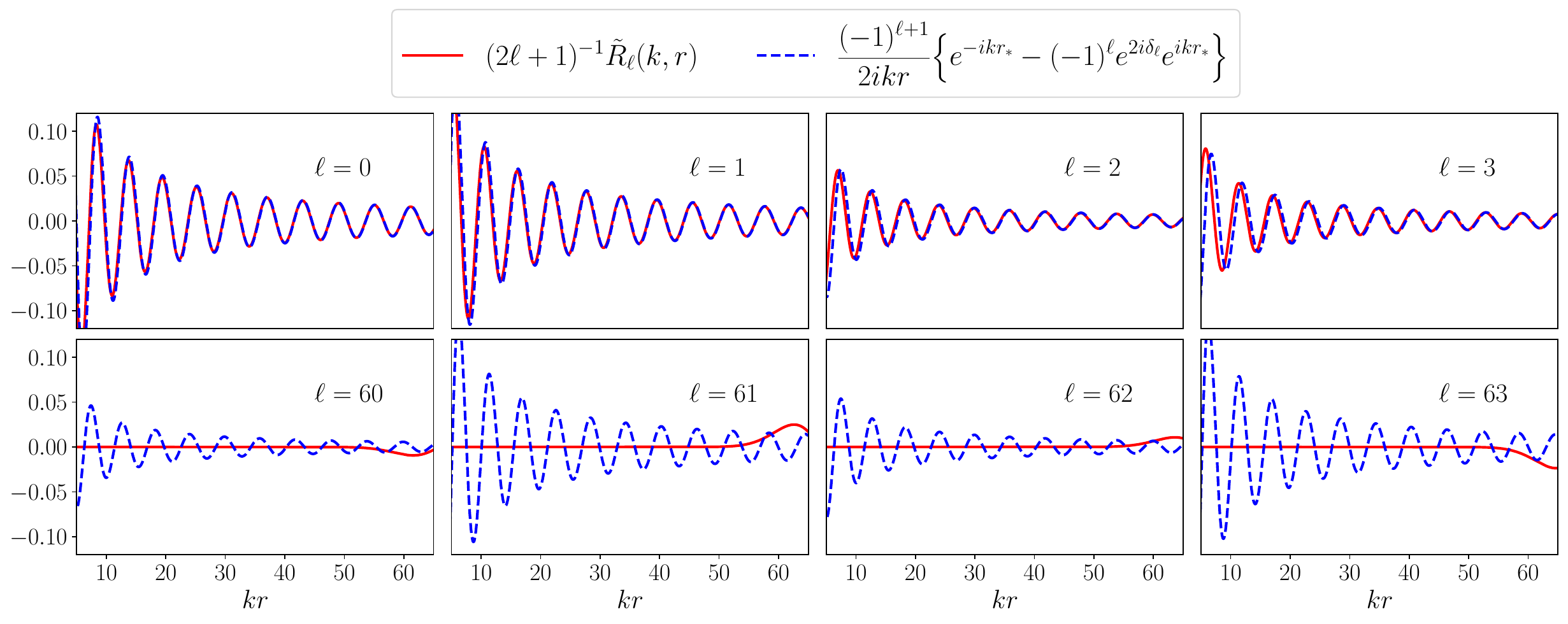}
    \caption{Comparison between the radial function $\tilde{R}_{\ell}(k,r)$, involving the Kummer hypergeometric function, and its large-$kr$ approximation, which is valid/invalid around $kr\sim60.0$ for the low/high-$\ell$ modes.}
    \label{fig:Kummer_asymptotic_expansione}
\end{figure}

\begin{figure}[H]
    \centering
    \begin{subfigure}[b]{0.45\textwidth}
        \includegraphics[width=\linewidth]{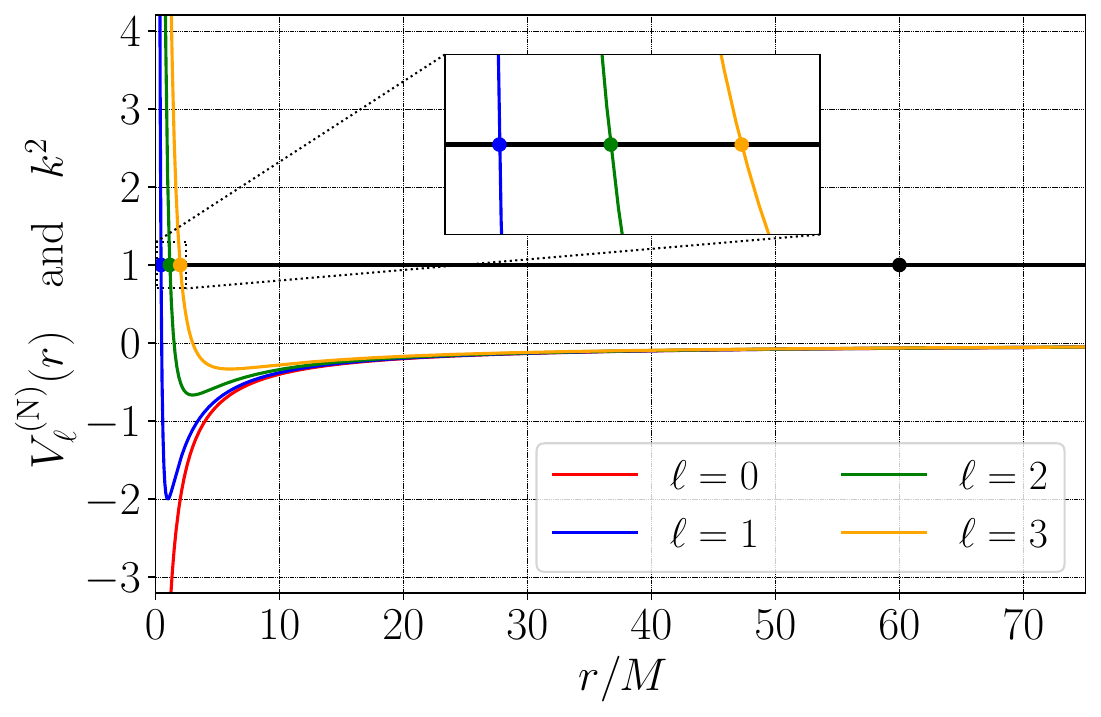}
        \caption{Low-$\ell$ modes.}
        \label{fig:Newton-potental-low-modes}
    \end{subfigure}
    \begin{subfigure}[b]{0.45\textwidth}
        \includegraphics[width=\linewidth]{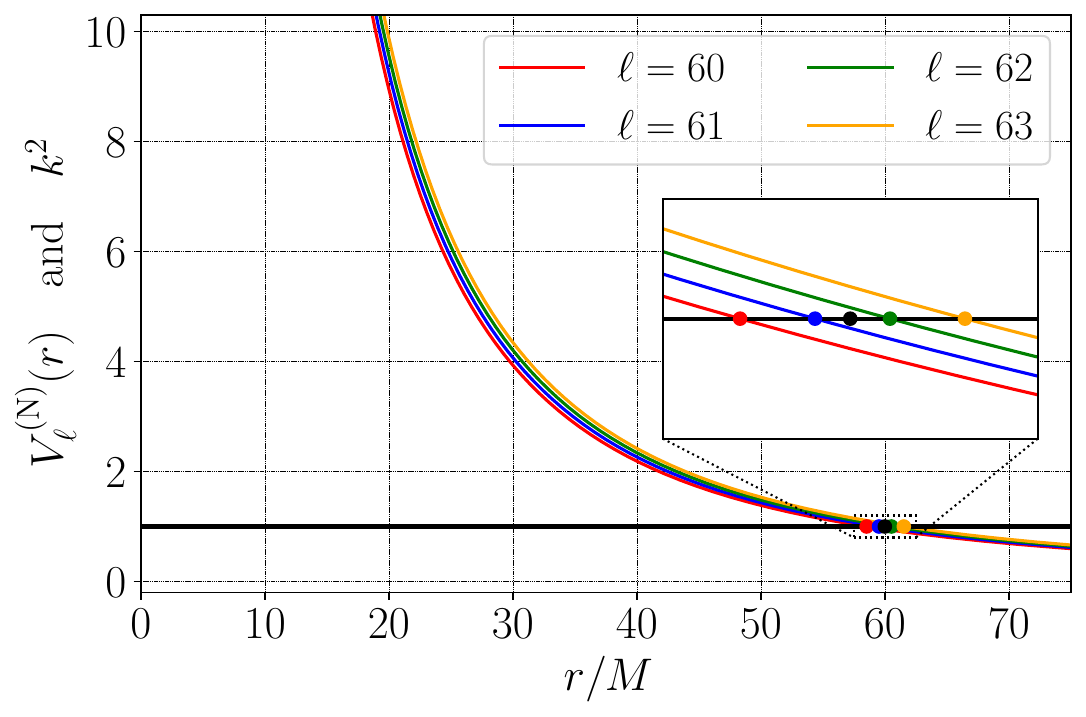}
        \caption{High-$\ell$ modes.}
        \label{fig:Newton-potental-high-modes}
    \end{subfigure}
\caption{Visual description of the location relation between the detector and turning point for the low/high-$\ell$ modes in the Newtonian scattering. In the weak-field approximation, the BH's event horizon is neglected, resulting in an infinitely high potential barrier when $r$ approaches $0$ (an infinitely deep potential well for $\ell=0$ mode). Such a scatterer completely reflects the scalar waves without absorption. There has to be a corresponding turning point for any frequency $k$ and $\ell\geqslant1$ modes.}
\label{fig:Newton-potental}
\end{figure}

\begin{figure}[H]
    \centering
    \includegraphics[width=1.0\linewidth]{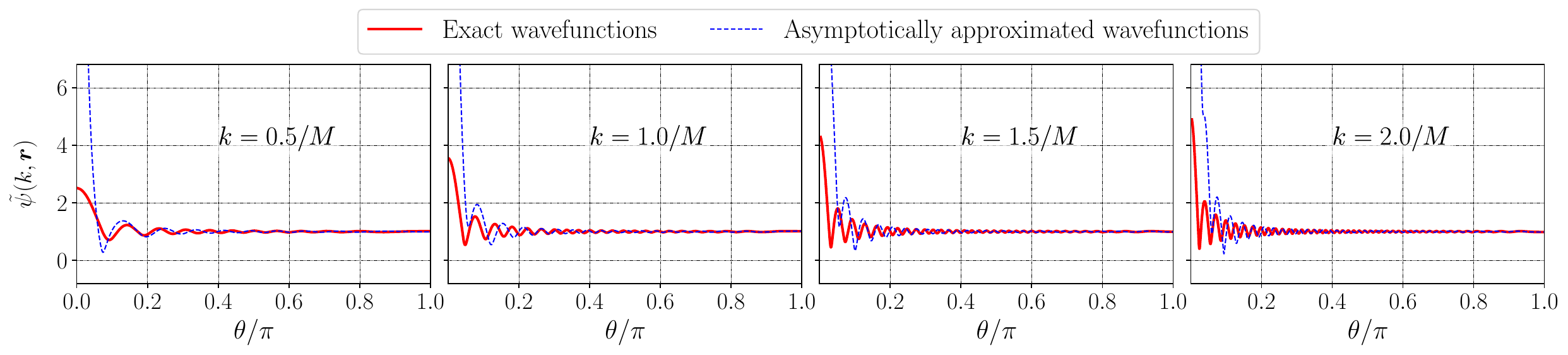}
    \caption{The wave functions calculated for $r=60.0M$, $\theta\in[0,\pi]$, and $k=\{0.5,1.0,1.5,2.0\}/M$. The red solid curves correspond to the exact wave function given by Eq.\,(\ref{eq-3:scattering-wave-function-spherical-harmonics-expansion}), and the blue dashed curves are evaluated from the asymptotic approximation (\ref{eq-3:wave-function-PW-asymptotic-expansion-reduced}), in which the angular function $\widehat{f}(\theta)$ has been calculated through the third-order SRM.}
    \label{fig:Newton-waveform-angular-distribution}
\end{figure}

\end{widetext}

Even though the asymptotic expansion leads to the divergence of the PWS, various regularization scheme have been developed to improve its convergence, such as SRM \cite{Yennie_1954,Stratton_2020,Dolan_2008b}, the Ces\`{a}ro summation \cite{Pijnenburg_2024b}, and the complex angular momentum method \cite{Andersson_1994,folacci_2019a,folacci_2019b} etc. As an example, in appendix \ref{app-A}, we review the basic principle of SRM, show the divergence of reduced series, and reproduce the results of the DCS $\dd\sigma/\dd\Omega=|f(\theta)|^2$.  Such regularization provides us an alternative approach to construct the scattering waveform, by approximating Eq.\,(\ref{eq-3:wave-function-PW-asymptotic-expansion}) as \cite{Chan_2025}
\begin{equation}
\label{eq-3:wave-function-PW-asymptotic-expansion-reduced}
\tilde{\psi}(k,\bm{r})\rightarrow e^{ikr\cos\theta}+\widehat{f}(\theta)\frac{e^{ikr_*}}{r}.
\end{equation}
$\widehat{f}(\theta)$ represents the regularized angular factor, which is computed through third-order SRM and is convergent for any off-axis position.

The numerical results given by Eq.\,(\ref{eq-3:wave-function-PW-asymptotic-expansion-reduced}) are shown in Fig.\,\ref{fig:Newton-waveform-angular-distribution}. Corresponding to this, we also calculate and display the exact results without asymptotic approximation, from Eq.\,(\ref{eq-3:scattering-wave-function-spherical-harmonics-expansion}). One finds that in the far-axis region, Eq.\,(\ref{eq-3:wave-function-PW-asymptotic-expansion-reduced}) captures the main characteristics of the exact wave function [see Eq.\,(\ref{eq-3:scattering-wave-function-spherical-harmonics-expansion})]. On the contrary, in the near-axis region, this approximation is invalid and is accompanied by a divergence, that is, the ``Poisson-spot divergence" mentioned in Section \ref{sec:intro}. In astrophysics, however, observers are often near the optical axis in a gravitational lensing system, highlighting the importance of constructing a scattering waveform that converges on the optical axis.

The on-axis divergence is mentioned twice in the previous discussion, specifically after Eq.\,(\ref{eq-3:Rutherford-formula}) and Eq.\,(\ref{eq-3:wave-function-PW-asymptotic-expansion-reduced}). The former is derived from the analytical parabolic-coordinate solution. At the same time, the latter, also referred to as the Poisson-spot divergence in this article, is found in the numerical calculations of $\widehat{f}(\theta)$. It is worth emphasizing that these two results are not independent of each other. The inappropriate asymptotic expansion of the wave function $\tilde{\psi}(k,r)$ near the optical axis resulted in a forward-scattering singularity, and the harmonics decomposition was forced to approximate this singularity. Therefore, the Poisson-spot divergence is only the manifestation of the forward-scattering singularity in the language of partial-wave analysis. This is the reason for the failure of regularization.

Before closing this subsection, we would like to emphasize the motivation of this work. For convenience, we suppose that the source is located at a radius $r_s$ and the observer is at a radius $r$. In realistic situations, these two distances are finite, and both the observer and the source stay distant from the scatterer, i.e., $r_s,r\gg M$. When applying the partial-wave method, the truncation is empirically taken as $\ell_{\max}\sim k\times\min\{r_s,r\}$ \cite{Kubota_2024,Santos_2025}, depending on the geometric scenario of the scattering system. Once one requires $r_s$ to be infinite, and adopts the plane-wave assumption to simplify the calculations, the truncation is completely determined by the radius $r$. Naturally, the existing empirical truncation requires $r$ to be a finite value all the time. Otherwise, for infinite $r$, the PWS does not converge at any finite $\ell_{\max}$, resulting in the divergent DCS. However, this scattering model, which simultaneously includes two infinities ($r\rightarrow\infty$ and $r_s\rightarrow\infty$), is employed in several previous works, leading to divergent PWS and Poisson spots. To overcome this issue, it seems necessary to avoid asymptotic expansion or the plane-wave assumption. This work first explores the formal case, in which the plane-wave assumption is adopted, while the asymptotic expansion is avoided to obtain physically reasonable results.

\section{\label{sec:RW}Black hole scattering}
\subsection{Computational procedure}
The last section investigated scalar scattering in weak gravity, where we approximated the gravitational potential as the Newtonian potential. In this section, we turn to the exact case, the scalar scattering by a Schwarzschild BH. The line element of the Schwarzschild metric in Schwarzschild coordinates $(t,r,\theta,\varphi)$ is \cite{weinberg}
\begin{equation}
\label{eq-4:Schwarzschild}
\dd s^2=-f(r)\dd t^2
+\frac{1}{f(r)}\dd r^2
+r^2(\dd\theta^2
+\sin^2\theta\dd\varphi^2),
\end{equation}
with $f(r)\equiv1-2M/r$, and $M$ being the BH mass. Starting from the massless KG equation (\ref{eq-3:KG}), transforming the wave function into Fourier space [see Eq.\,(\ref{eq-3-wave-function-Fourier})], and then taking harmonic decomposition [see Eq.\,(\ref{eq-3:scattering-wave-function-spherical-harmonics-expansion})], we finally arrive at the spin-0 RW equation \cite{Maggiore_2018}
\begin{equation}
\label{eq-4:Regge-Wheeler-equation}
\left[\frac{\dd^2}{\dd r_*^2}+k^2-V_{\ell}(r)\right]u_{\ell}(k,r)=0.
\end{equation}
The radial function is defined as $u_{\ell}(k,r)\equiv r\tilde{R}_{\ell}(k,r)$, with the tortoise coordinate being 
\begin{equation}
\label{eq-4-r-star}
r_*\equiv r+2M\ln(r/2M-1).
\end{equation}
The RW potential $V_{\ell}(r)$ for the spin-0 perturbation is
\begin{equation}
V_{\ell}(r)=\left(1-\frac{2M}{r}\right)\left\{\frac{\ell(\ell+1)}{r^2}+\frac{2M}{r^3}\right\}.
\end{equation}
A main difference between RW and Newtonian potentials (\ref{eq-3-Newton-potential}) is that the RW potential tends to $0$ as $r$ approaches $2M$, due to the absorption of scalar waves by the BH's event horizon. 

For arbitrary angular quantum number $\ell$, the ``in" solution to Eq.\,(\ref{eq-4:Regge-Wheeler-equation}) is approximated as
\begin{equation}
\label{eq-4:outer-boundary-condition}
u_{\ell,{\rm in}}(k,r)\rightarrow
A_{\ell}e^{-ikr_*}+B_{\ell}e^{ikr_*},
\end{equation}
for $kr\gg\ell$, and
\begin{equation}
\label{eq-4:inner-boundary-condition}
u_{\ell,{\rm in}}(k,r)\rightarrow
e^{-ikr_*},
\end{equation}
for $r\rightarrow2M$ or $r_*\rightarrow-\infty$. To solve Eq.\,(\ref{eq-4:Regge-Wheeler-equation}) numerically, one first sets the inner boundary condition as Eq.\,(\ref{eq-4:inner-boundary-condition}), and then evolves $u_{\ell}$ to the outer boundary $r_{\max}$, which is required to be sufficiently distant from the turning point. Matching the numerical result with the outer boundary condition (\ref{eq-4:outer-boundary-condition}) determines the incident coefficient $A_{\ell}$ and reflection coefficient $B_{\ell}$. Let the incident coefficient $A_{\ell}$ to be consistent with incident plane wave (\ref{eq-2:expansion-coefficients-asymptotic-expansion}), one gets
\begin{equation}
A_{\ell}=(-1)^{\ell+1}\frac{(2\ell+1)}{2ik}.
\end{equation}
The $\ell$-order partial wave is
\begin{equation}
\tilde{R}_{\ell}(k,r)=\frac{u_{\ell}(k,r)}{r}=\frac{A_{\ell}(k)}{r}\widehat{u}_{\ell,\text{in}}(k,r),
\end{equation}
and the full frequency-domain scattered scalar wave, defined in Eq.\,(\ref{eq-3:scattering-wave-function-spherical-harmonics-expansion}), is given by 
\begin{equation}
\label{eq-4:scattering-wave-function}
\begin{aligned}
\tilde{\psi}(k,\bm{r})
=\frac{1}{2ikr}\sum_{\ell=0}^{\infty}
(-1)^{\ell+1}(2\ell+1)\widehat{u}_{\ell,\text{in}}(k,r)
\mathrm{P}_{\ell}(\cos\theta),
\end{aligned}
\end{equation}
where $\widehat{u}_{\ell,\text{in}}(k,r)=A_{\ell}^{-1}u_{\ell,\text{in}}(k,r)$ is the regularized radial function, which obeys the following outer boundary condition,
\begin{equation}
\label{eq-4-outer-boundary-condition}
\widehat{u}_{\ell,\text{in}}(k,r)\rightarrow
e^{-ikr_*}-(-1)^{\ell}e^{2i\delta_{\ell}}e^{ikr_*},
\end{equation}
for $kr\gg\ell$, and inner boundary condition
\begin{equation}
\label{eq-4-inner-boundary-condition}
\widehat{u}_{\ell,\text{in}}(k,r\rightarrow2M)\rightarrow
A_{\ell}^{-1}e^{-ikr_*},
\end{equation}
The phase shift $e^{2i\delta_{\ell}}$ has been defined as \cite{Dolan_2008b}
\begin{equation}
\label{eq-4-phase-shift}
e^{2i\delta_{\ell}}=-(-1)^{\ell}\frac{B_{\ell}}{A_{\ell}}.
\end{equation}

From the phase shift (\ref{eq-4-phase-shift}), one can calculate $f(\theta)$ from Eq.\,(\ref{eq-3:wave-function-PW-asymptotic-expansion-scattered-wave}) and DCS. For the far-axis region, the waveform is approximated by Eq.\,(\ref{eq-3:wave-function-PW-asymptotic-expansion-reduced}), where $f(\theta)$ is regularized through SRM. It should be emphasized again that in this work, we present the rigorous computation of the scattered wave by avoiding the asymptotic expansion. In our computation, the observer is always taken at a finite radius $r$, and, therefore, the PWS is truncated at $\ell_{\max}\sim kr$ empirically. For all partial waves with $\ell\lesssim\ell_{\max}$, the outer boundary $r_{\max}$ is always set to be sufficiently distant to the turning point $\hat{r}_{\ell}$, where the approximation condition (\ref{eq-4:outer-boundary-condition}) is applicable and the regularization factor $A_{\ell}$ is correctly determined.

\subsection{Numerical results}
In this subsection, Eq.\,(\ref{eq-4:scattering-wave-function}) is numerically computed, and the corresponding results are plotted in Fig.\,\ref{fig:scattering-wave-RW} for various frequencies. The wave fields are quite similar to those in Newtonian scattering. In the negative-$z$ region, the scalar field is roughly a plane wave, with slight wavefront distortion due to the long-range interaction. In the positive-$z$ but far-axis region, the fields behave as the superposition of distorted plane waves and outgoing spherical waves. In the vicinity of the optical axis, strong interference patterns and Poisson spots appear. 

Fig.\,\ref{fig:RW-convergence} presents the convergence of the PWS of Eq.\,(\ref{eq-4:scattering-wave-function}). Not surprisingly, for all points under investigation, including one on the optical axis, the PWS converge and are truncated at a certain $\ell_{\max}$ value. Unfortunately, BH scattering does not have an analytical solution like Eq.\,(\ref{eq-3:scattering-wave-function-paraboloidal}), so there are no results compared to PWS shown in Fig.\,\ref{fig:RW-convergence}.

Fig.\,\ref{fig:RW_asymptotic_expansione} compares the exact radial function $\tilde{R}_{\ell}(k,r)$ and its large-$kr$ expansion. As mentioned before, such an approximation is only valid for the low-$\ell$ (e.g., $\ell\ll kr$) partial waves, where the observer is sufficiently distant from the turning point, i.e., $r\gg\hat{r}_{\ell}$. But this expansion fails to approximate the high-$\ell$ (e.g., $\ell\sim kr$) modes, where the observer is roughly located at the turning point. The above discussion is visually shown in Fig.\,\ref{fig:RW-potental}, which illustrates the positional relationship between the observer, potential function, and turning point. 

In Fig.\,\ref{fig:RW-waveform-angular-distribution}, we compare the exact and asymptotically-approximated wave functions evaluated in fixed radius and various polar angles. The approximated wave function is given by Eq.\,(\ref{eq-3:wave-function-PW-asymptotic-expansion-reduced}), with the phase shift being given by Eq.\,(\ref{eq-4-phase-shift}). The asymptotic expansion captures the dominant feature of exact results in the far-axis region, which, unfortunately, is imprecise. On the contrary, the asymptotic expansion significantly deviates from the rigorous result in the near-axis region, e.g., $\theta\lesssim\pi/5$, ultimately leading to the divergence of Poisson spots. 

\begin{widetext}

\begin{figure}[H]
    \centering
    \begin{subfigure}[b]{0.24\textwidth}
        \includegraphics[width=\linewidth]{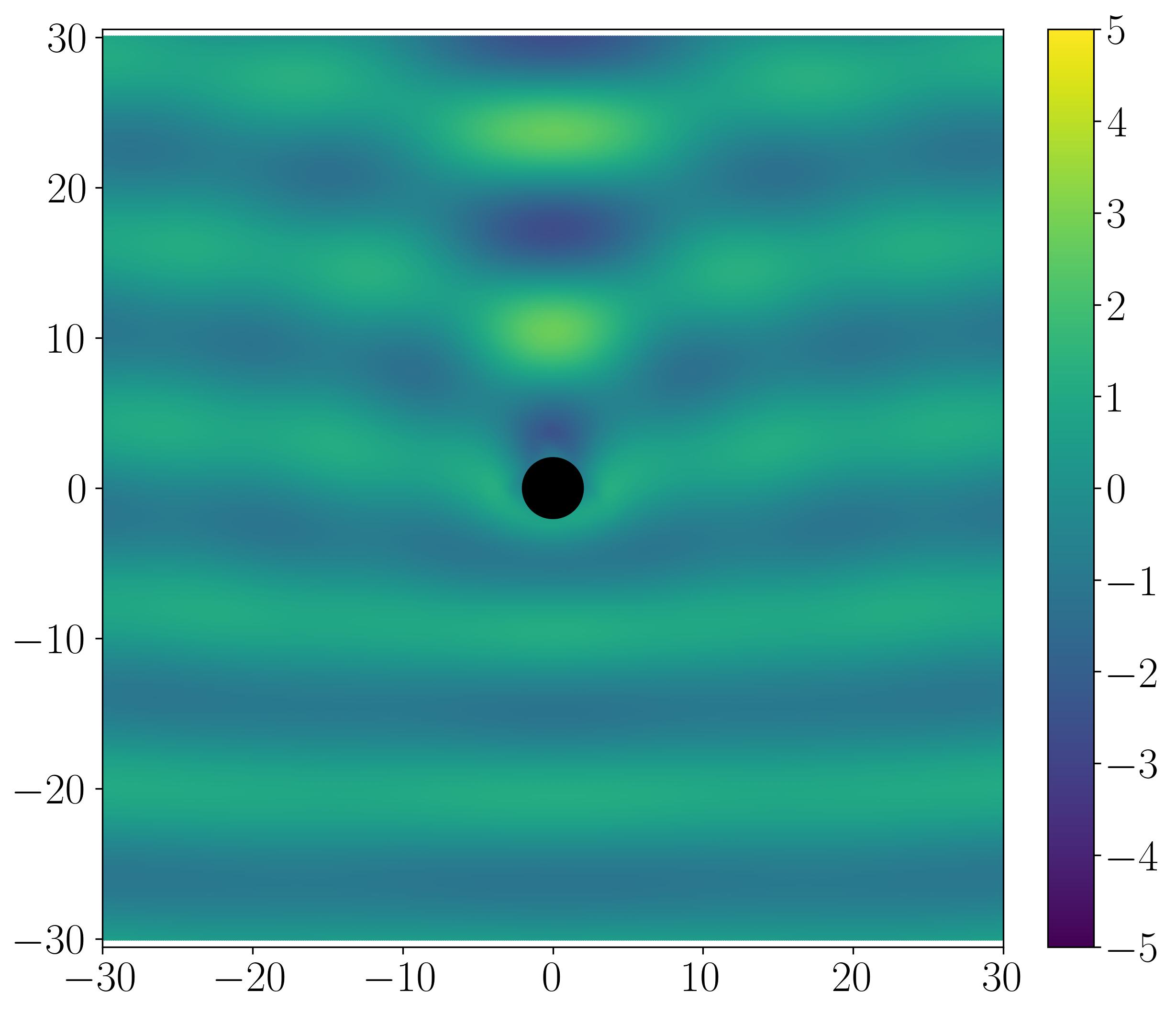}
        \caption{$\mathrm{Re}\,\Big\{\tilde{\psi}(k=0.5/M,\bm{r})\Big\}$}
    \end{subfigure}
    \begin{subfigure}[b]{0.24\textwidth}
        \includegraphics[width=\linewidth]{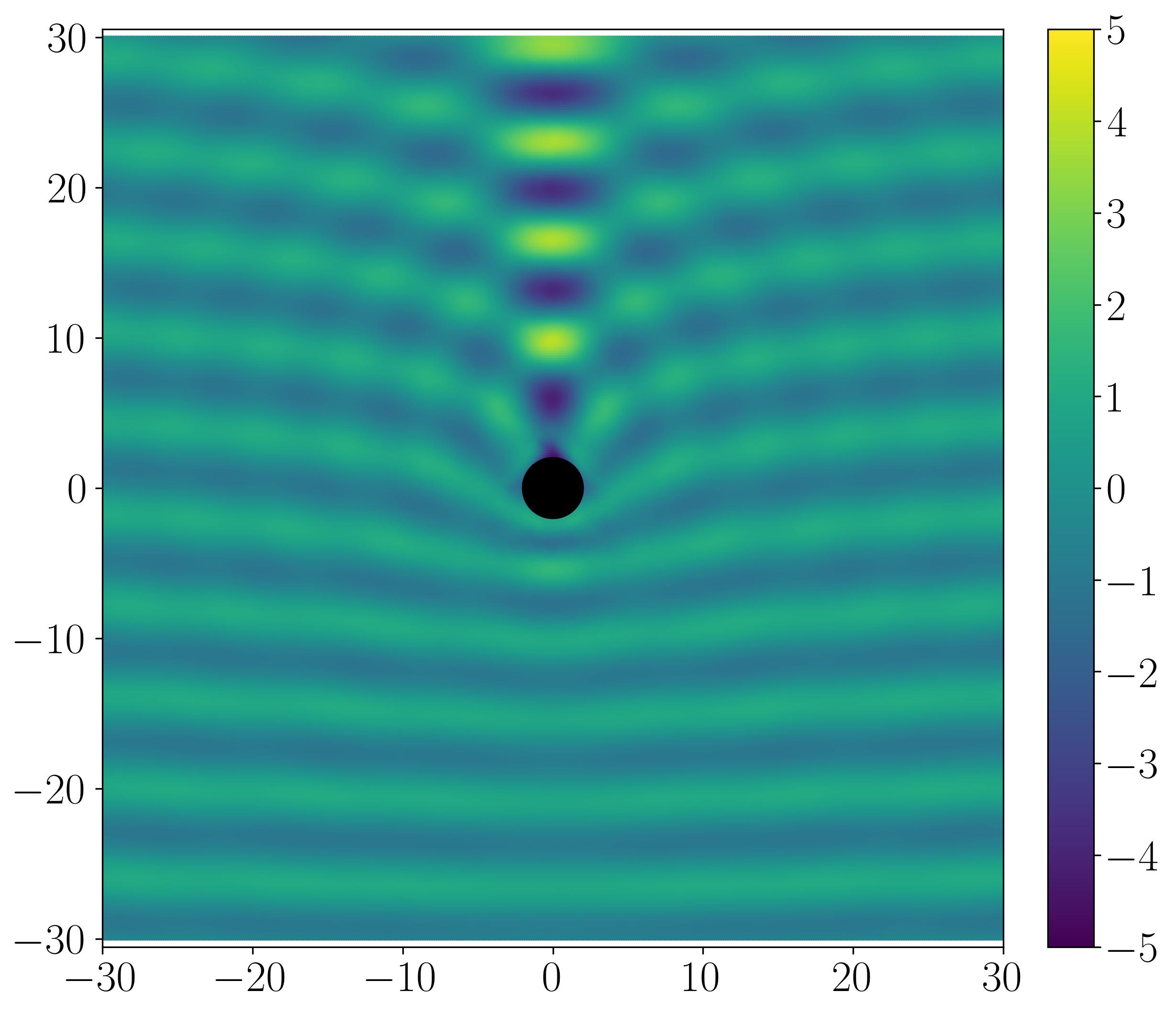}
        \caption{$\mathrm{Re}\,\Big\{\tilde{\psi}(k=1.0/M,\bm{r})\Big\}$}
    \end{subfigure}
    \begin{subfigure}[b]{0.24\textwidth}
        \includegraphics[width=\linewidth]{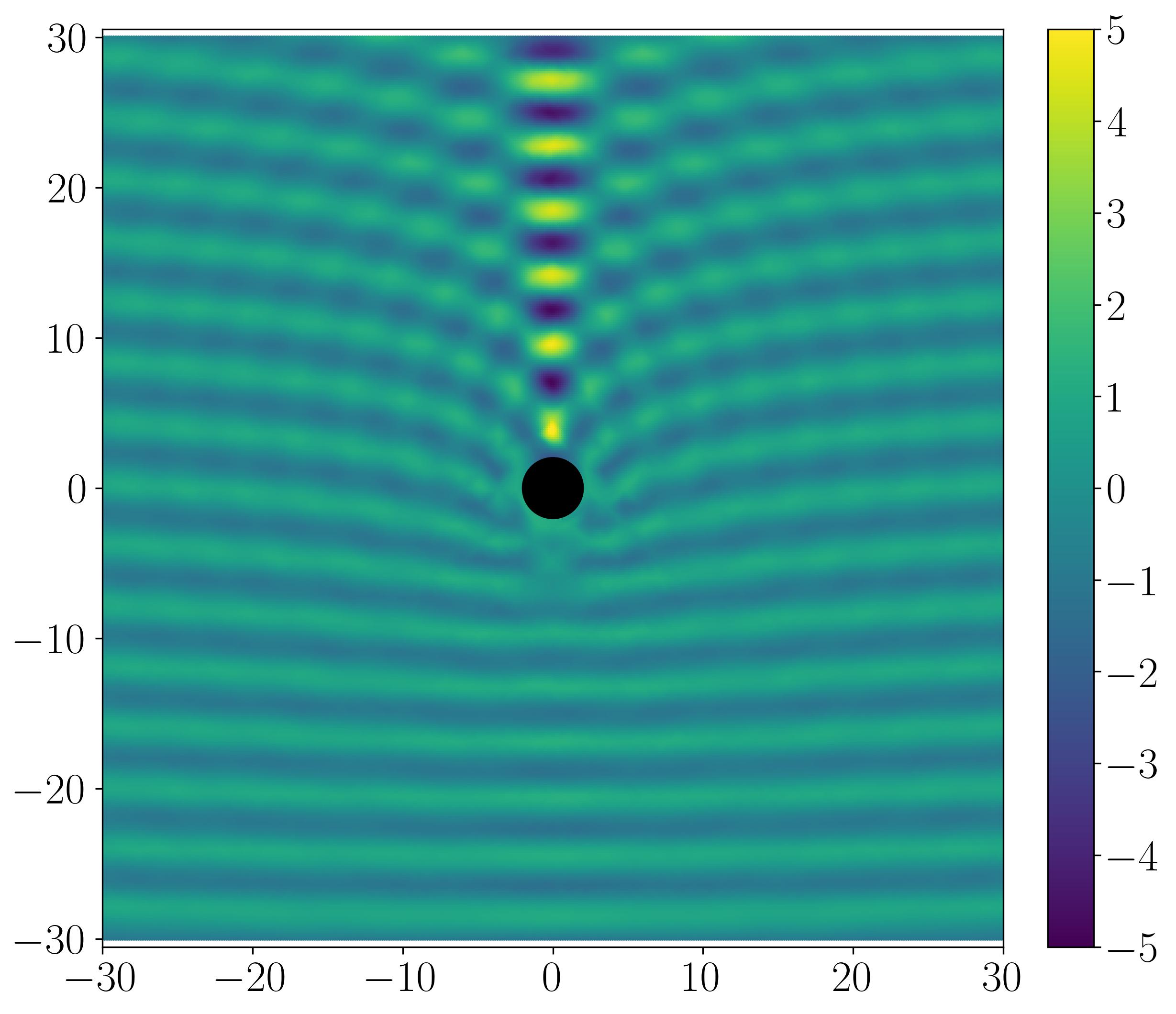}
        \caption{$\mathrm{Re}\,\Big\{\tilde{\psi}(k=1.5/M,\bm{r})\Big\}$}
    \end{subfigure}
    \begin{subfigure}[b]{0.24\textwidth}
        \includegraphics[width=\linewidth]{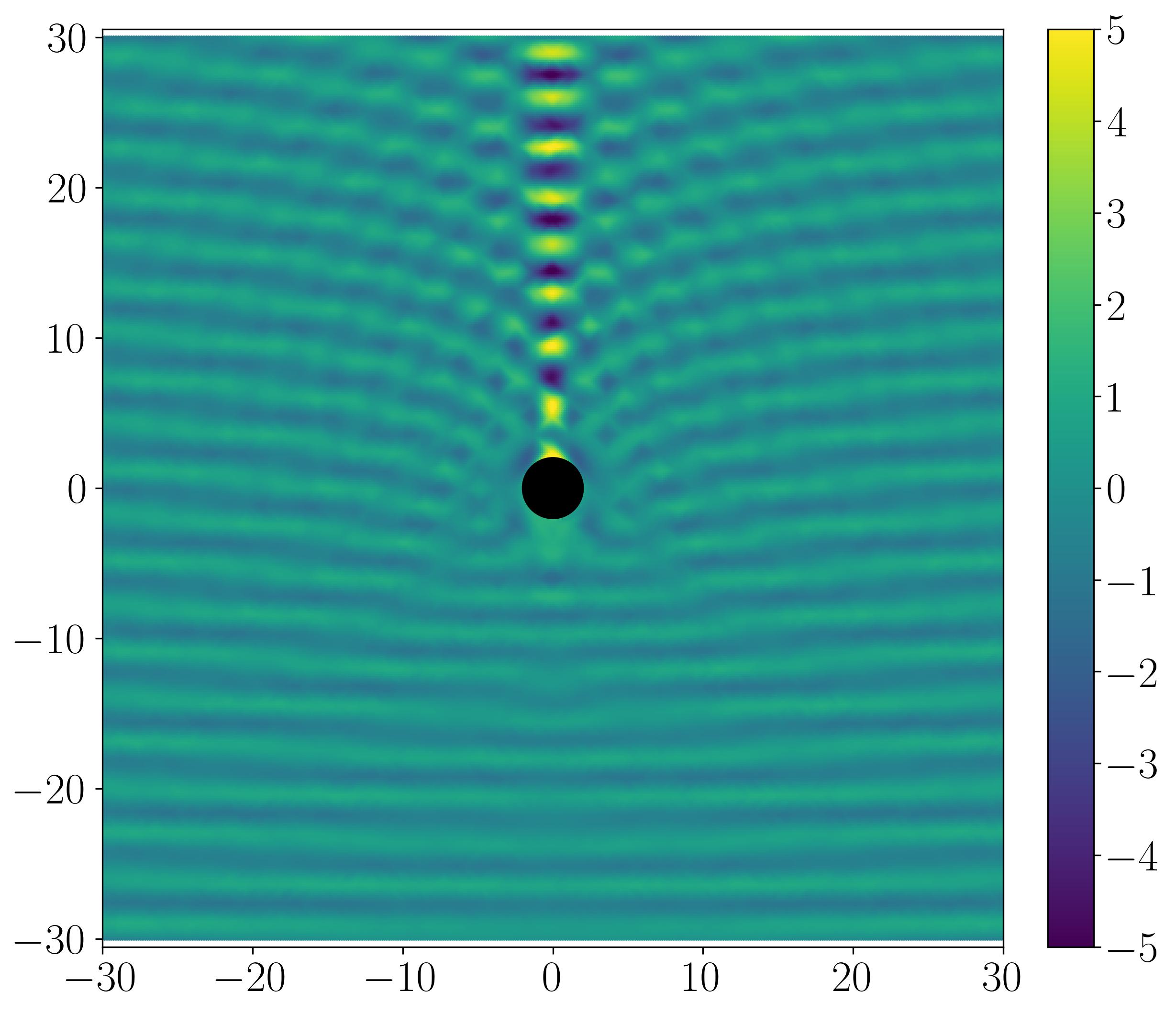}
        \caption{$\mathrm{Re}\,\Big\{\tilde{\psi}(k=2.0/M,\bm{r})\Big\}$}
    \end{subfigure}

    \begin{subfigure}[b]{0.24\textwidth}
        \includegraphics[width=\linewidth]{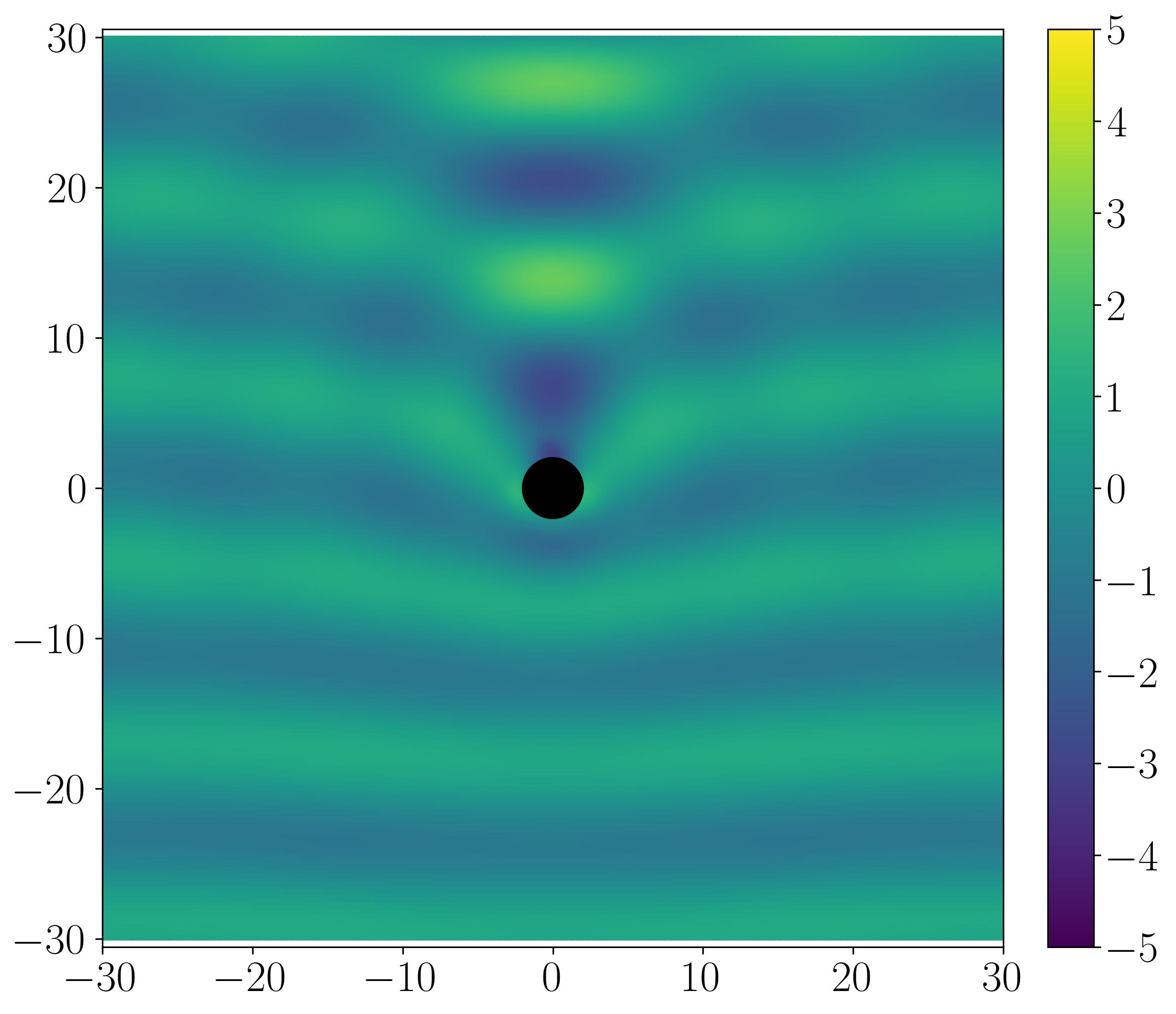}
        \caption{$\mathrm{Im}\,\Big\{\tilde{\psi}(k=0.5/M,\bm{r})\Big\}$}
    \end{subfigure}
    \begin{subfigure}[b]{0.24\textwidth}
        \includegraphics[width=\linewidth]{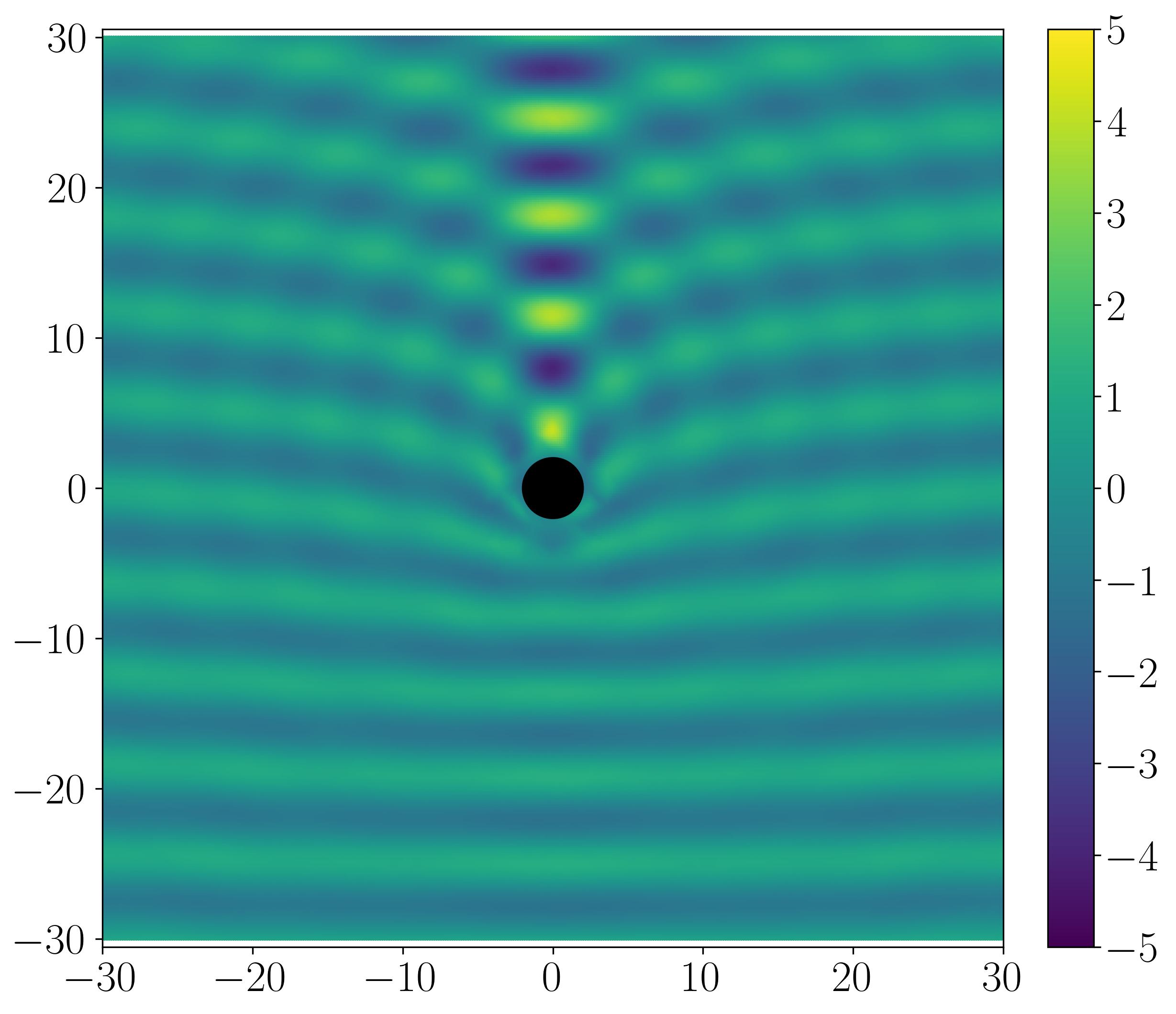}
        \caption{$\mathrm{Im}\,\Big\{\tilde{\psi}(k=1.0/M,\bm{r})\Big\}$}
    \end{subfigure}
    \begin{subfigure}[b]{0.24\textwidth}
        \includegraphics[width=\linewidth]{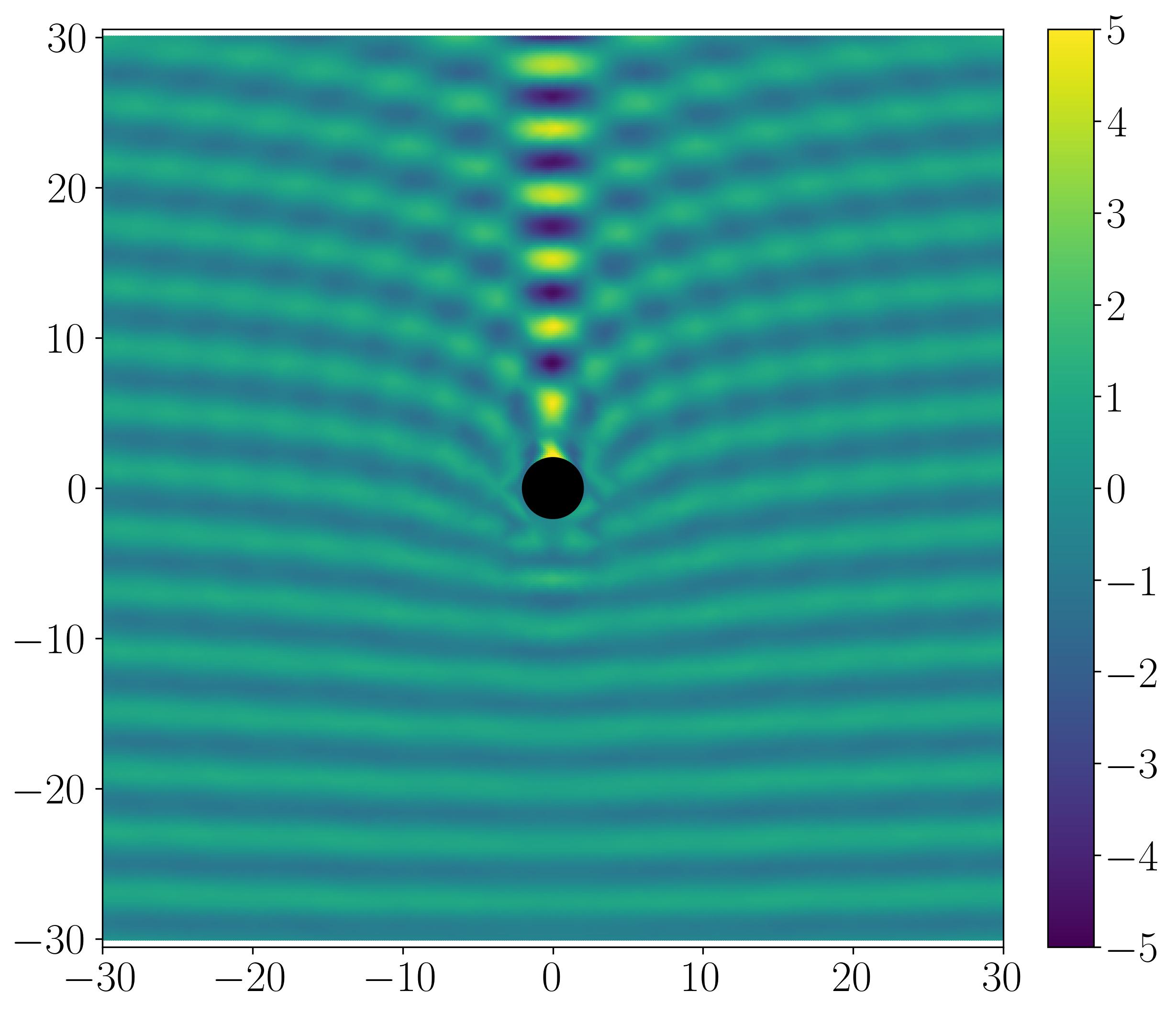}
        \caption{$\mathrm{Im}\,\Big\{\tilde{\psi}(k=1.5/M,\bm{r})\Big\}$}
    \end{subfigure}
    \begin{subfigure}[b]{0.24\textwidth}
        \includegraphics[width=\linewidth]{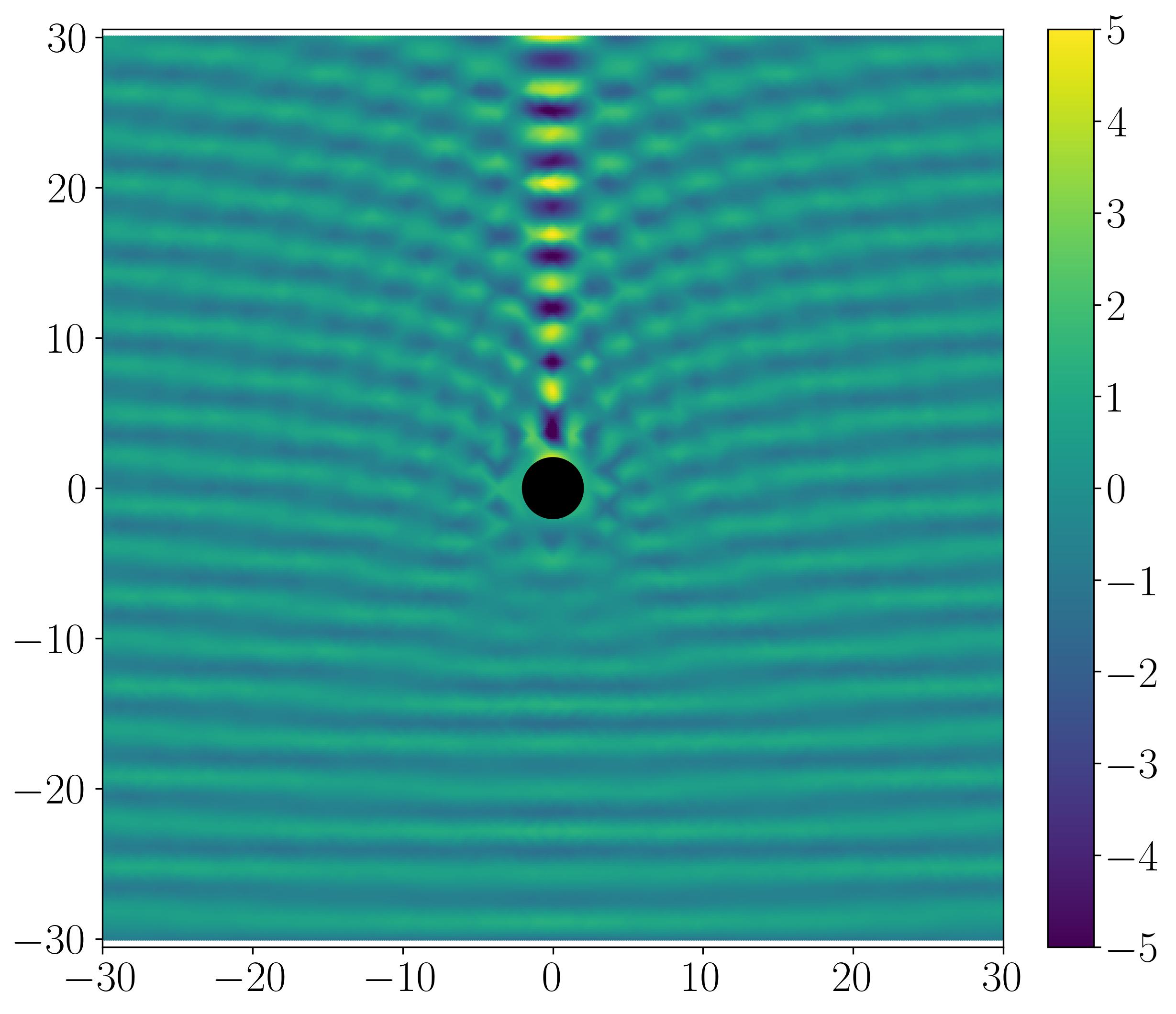}
        \caption{$\mathrm{Im}\,\Big\{\tilde{\psi}(k=2.0/M,\bm{r})\Big\}$}
    \end{subfigure}
    \caption{Scattered monochromatic scalar wave fields calculated through Eq.\,(\ref{eq-4:scattering-wave-function}) with $k=\{0.5,1.0,1.5,2.0\}/M$. The black spots, with radius $2M$, represent the central BH.}
    \label{fig:scattering-wave-RW}
\end{figure}

\begin{figure}[htbp]
    \centering
    \begin{subfigure}[b]{0.49\textwidth}
        \includegraphics[width=\linewidth]{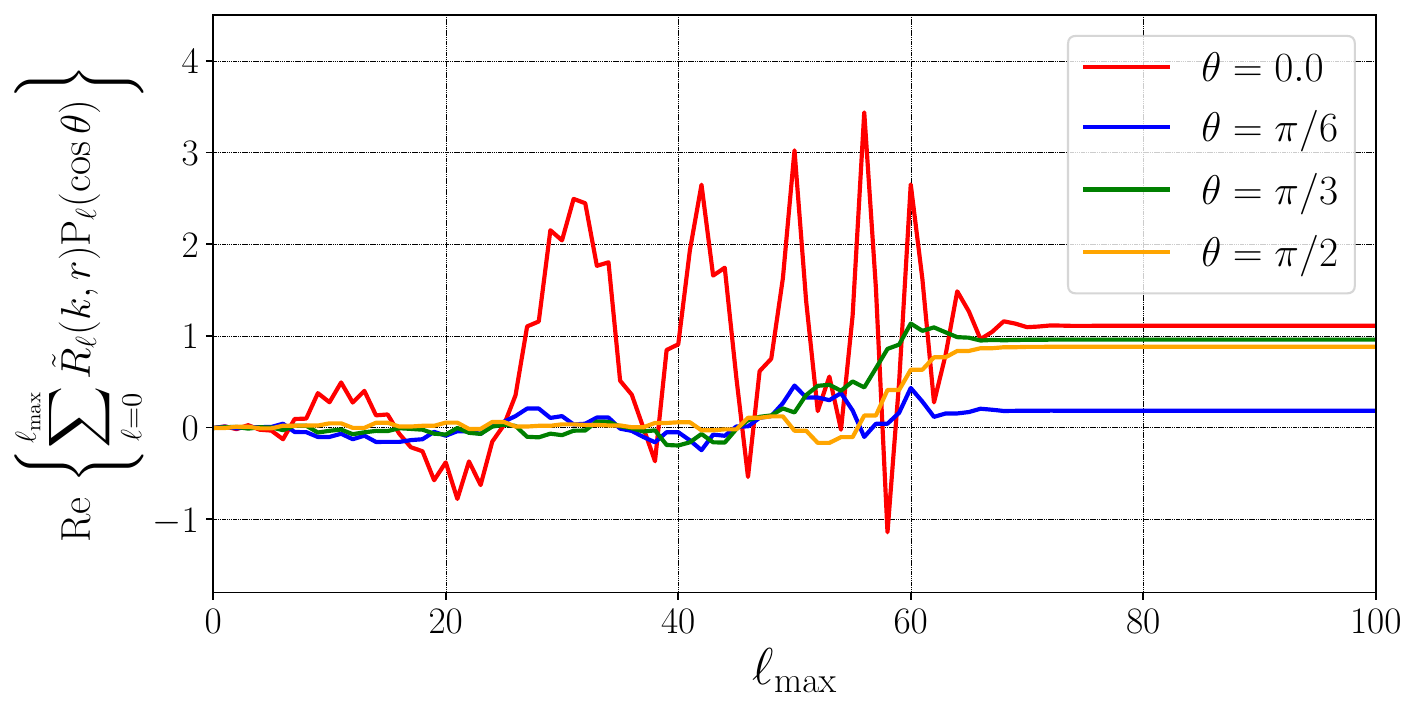}
    \end{subfigure}
    \begin{subfigure}[b]{0.49\textwidth}
        \includegraphics[width=\linewidth]{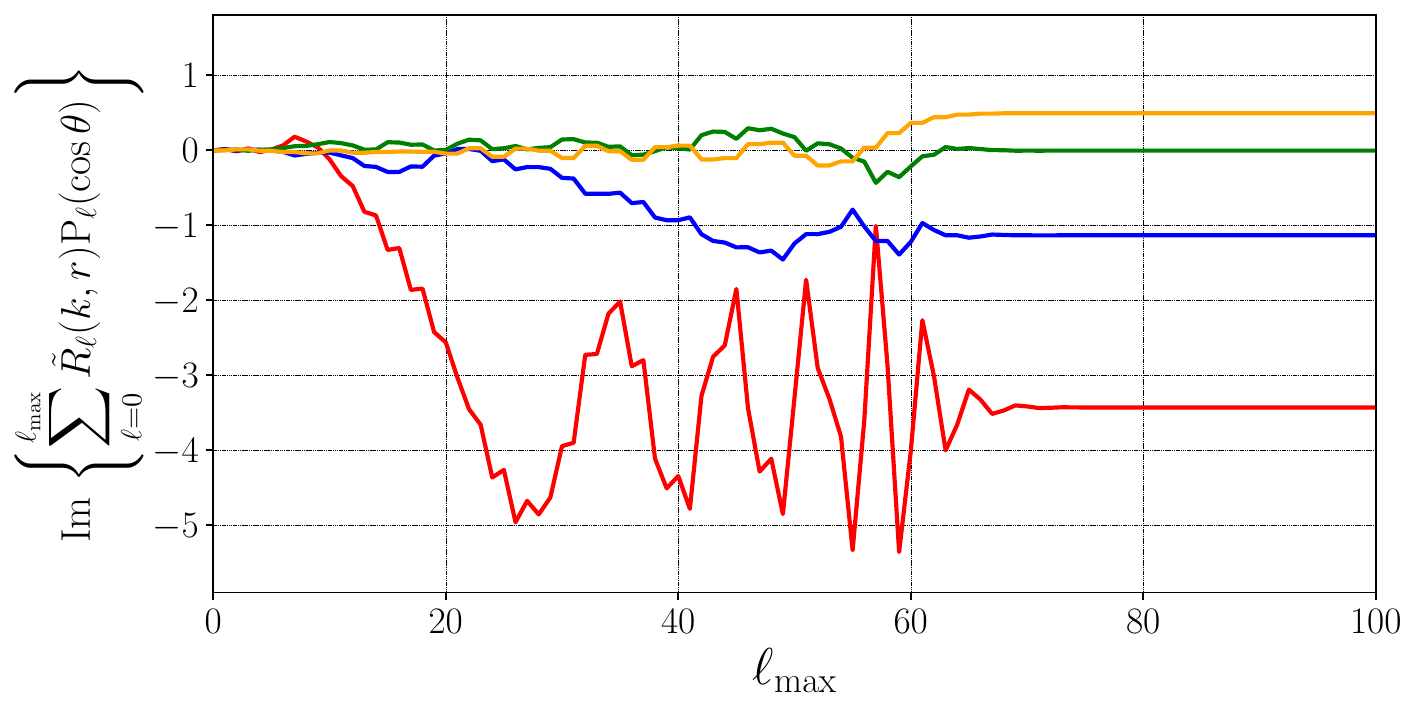}
    \end{subfigure}
\caption{The convergence of Eq.\,(\ref{eq-4:scattering-wave-function}). The observer is located at $r=60.0M$ and $\theta=\{0,\pi/6,\pi/3,\pi/2\}$. At all sampled points, the PWS are convergent and truncated at around $\ell_{\max}\sim kr$, including on the optical axis. }
\label{fig:RW-convergence}
\end{figure}

\begin{figure}[H]
    \centering
    \includegraphics[width=0.90\textwidth]{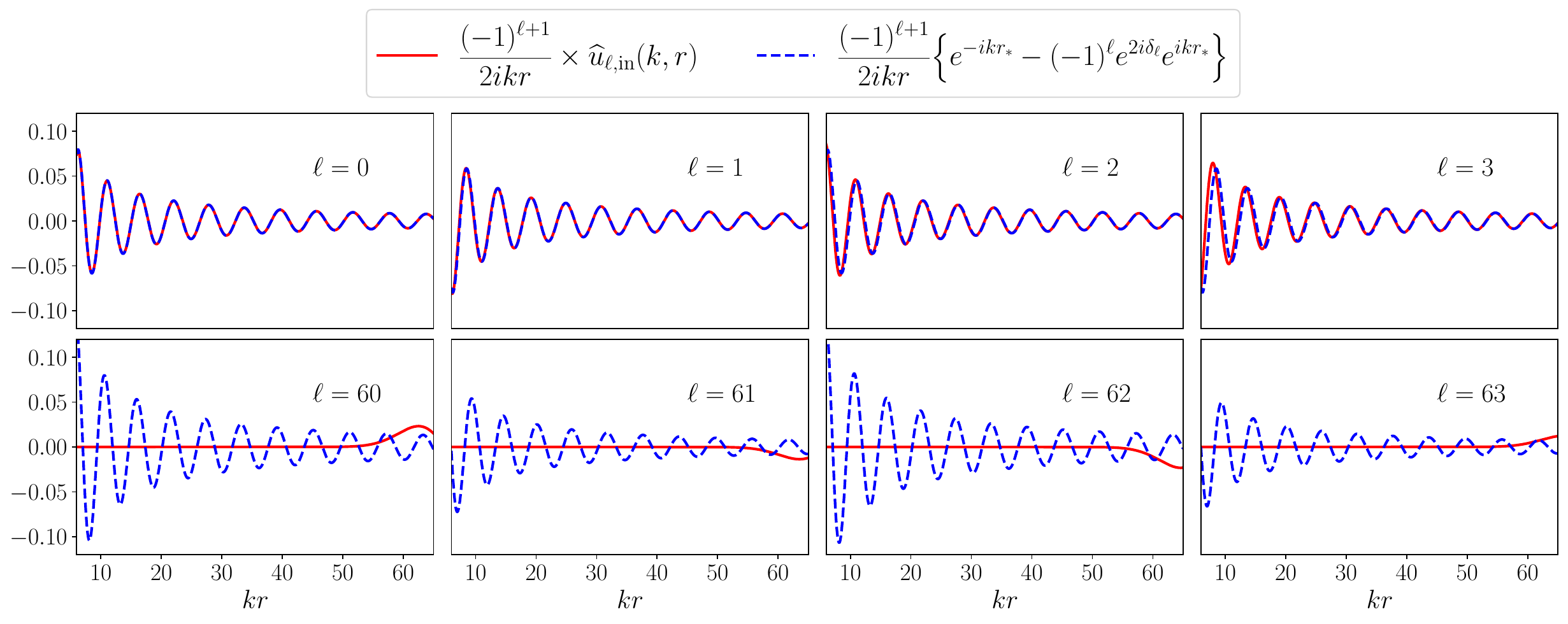}
    \caption{The comparison between the radial function $\tilde{R}_{\ell}(k,r)$ and its large-$kr$ approximation, the latter is valid/invalid around $kr\sim60.0$ for the low/high-$\ell$ partial waves.}
    \label{fig:RW_asymptotic_expansione}
\end{figure}

\begin{figure}[H]
    \centering
    \begin{subfigure}[b]{0.45\textwidth}
        \includegraphics[width=\linewidth]{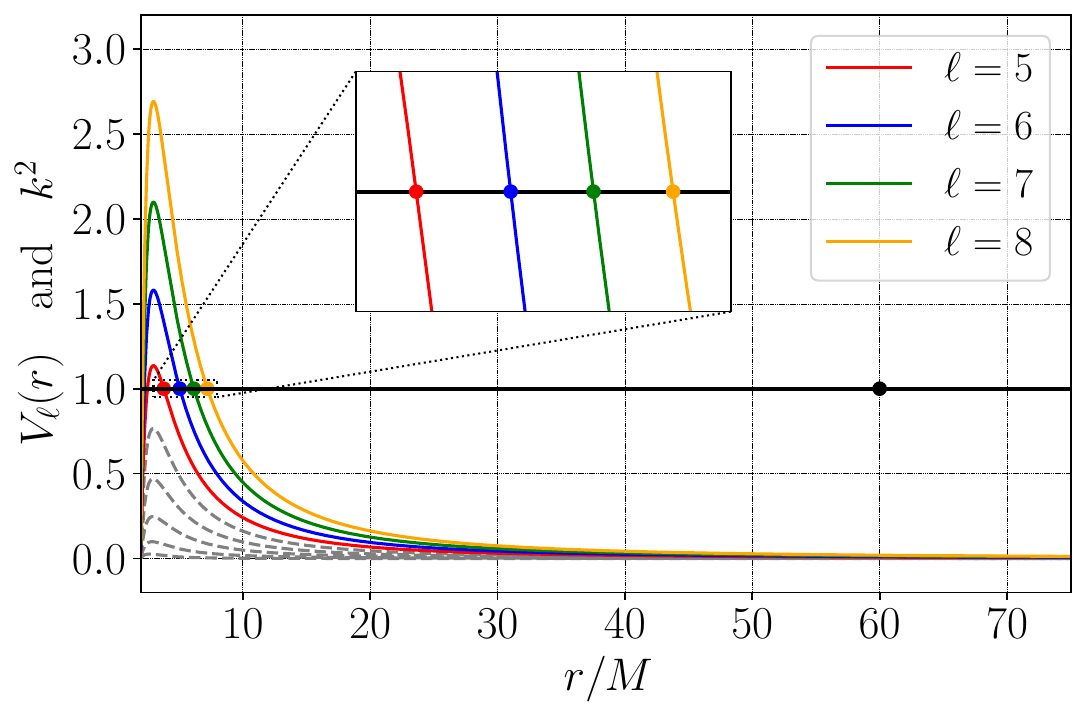}
        \caption{Low-$\ell$ modes.}
        \label{fig:RW-potental-low-modes}
    \end{subfigure}
    \begin{subfigure}[b]{0.45\textwidth}
        \includegraphics[width=\linewidth]{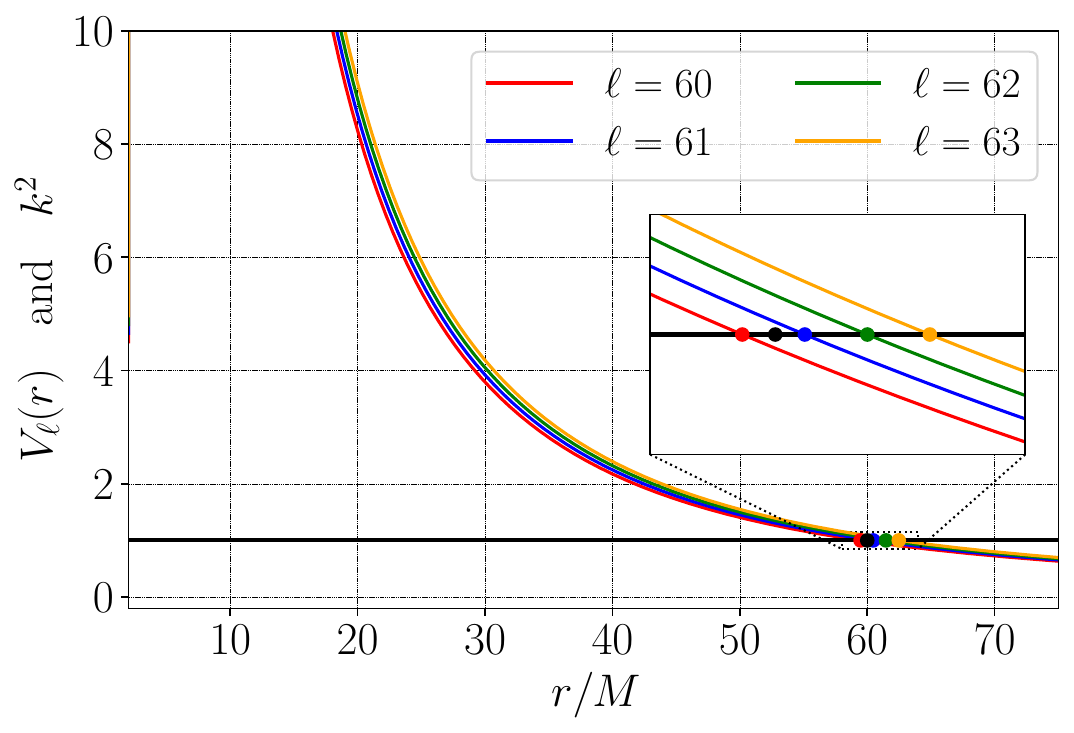}
        \caption{High-$\ell$ modes.}
        \label{fig:RW-potental-high-modes}
    \end{subfigure}
\caption{Visual description of the location relation between the detector turning point for the low/high-$\ell$ partial waves in the BH scattering. In the BH scattering model, where the event horizon has been taken into account, the potential is a barrier for arbitrary $\ell$, and possesses a peak at about $r\sim3M$. There is no turning point when $k^2>V_{\rm peak}$, and there have to be two turning points when $k^2<V_{\rm peak}$. For the scattering process, we focus on the outer one. In our example, where $k=1.0/M$, the first turning point appears in $\ell=5$ mode.}
\label{fig:RW-potental}
\end{figure}

\begin{figure}[H]
    \centering
    \includegraphics[width=1.0\linewidth]{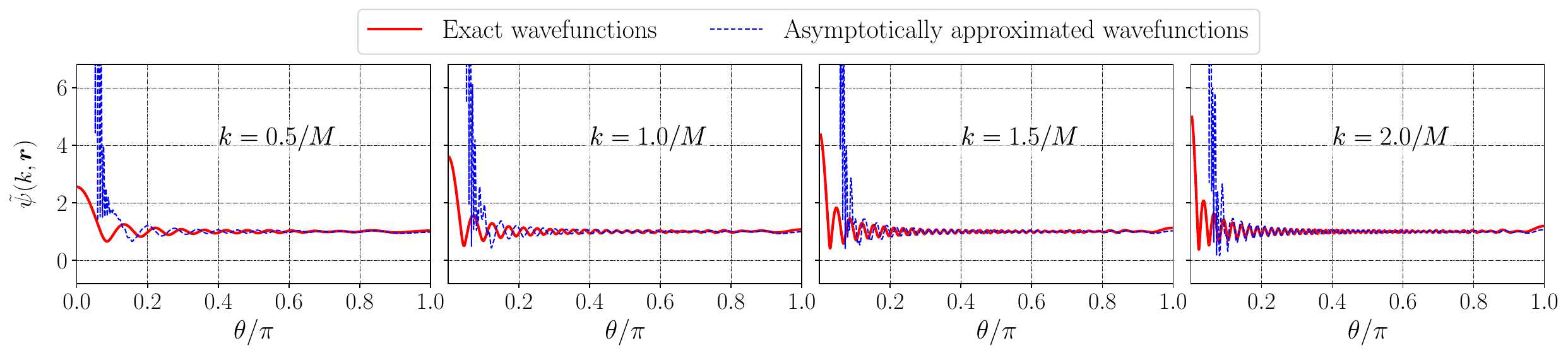}
    \caption{The wave functions calculated for $r=60.0M$, $\theta\in[0,\pi]$, and $k=\{0.5,1.0,1.5,2.0\}/M$. The red solid curves correspond to Eq.\,(\ref{eq-4:scattering-wave-function}), and the blue dashed curves correspond Eq.\,(\ref{eq-3:wave-function-PW-asymptotic-expansion-reduced}). The phase shift is given by Eq.\,(\ref{eq-4-phase-shift}) and the angular factor $\widehat{f}(\theta)$ is calculated through the third-order SRM.}
    \label{fig:RW-waveform-angular-distribution}
\end{figure}

\end{widetext}

\subsection{\label{subsec:transmission}Scalar scattering as the high-frequency approximation of gravitational wave scattering}
The KG equation (\ref{eq-3:KG}) describes not only the massless scalar fields, but also the high-frequency GWs on curved backgrounds. Starting from the LEFE and through the eikonal expansion, the metric of high-frequency GWs is approximated as a spin-0 scalar field without polarization. The reader can find more details in Refs.\,\cite{mtw,Isaacson_1968,Hou_2019} or Appendix \ref{app-B}. 

Without ambiguity, we still represent the amplitude of GWs as $\tilde{\psi}(k,\bm{r})$. In the frequency domain, the ratio between lensed and unlensed waveforms is defined as the transmission factor, denoted by $F(k)$. For a point-like and weak-gravity lens object with mass $M$, $F(k)$ is solved from Eq.\,(\ref{eq-3:KG-weak-field}) as \cite{Takahashi_2003}
\begin{equation}
\label{eq-4-Kirchhoff}
\begin{aligned}
F(k)&=e^{\pi\gamma/2}\left(-\gamma\right)^{-i\gamma}
\Gamma(1+i\gamma)\\
&\qquad\times{_1F_1}\left\{-i\gamma,1;-i\gamma\left(\frac{\xi}{\xi_0}\right)^2\right\},
\end{aligned}
\end{equation}
through the diffraction integral in the near-axis region. Following the conventional notation, we denote the observer, wave source, and lens/scatterer as ``O", ``S", and ``L", respectively. The plane perpendicular to ``LO" and containing ``L" is called the lens plane. The angular coordinate of source on the lens plane as $\xi\equiv\angle\text{LOS}$. Similarly, the plane perpendicular to ``SO" and containing ``S" is called the source plane. As a habit, one introduces the Einstein angle $\xi_0\equiv(4MD_{LS}/D_{L}D_{S})^{1/2}$ as the normalization of the angular coordinate, where $D_L$, $D_S$, and $D_{LS}$ are the distances from observer to lens plane, from observer to source plane, and from source plane to lens plane, respectively. When considering the planar incident wave, we let $D_{L}=r$ and $D_{S}, D_{LS}\rightarrow\infty$ in Eq.\,(\ref{eq-4-Kirchhoff}). Therefore, the Einstein radius is $\xi_0=(4M/r)^{1/2}$, and the dimensionless source coordinate is $\xi/\xi_0=(1/2)(r/M)^{1/2}\tan\theta$, where $(r,\theta)$ is Schwartzschild coordinate of observer.

\begin{widetext}

\begin{figure}[H]
    \centering
    \includegraphics[width=\linewidth]{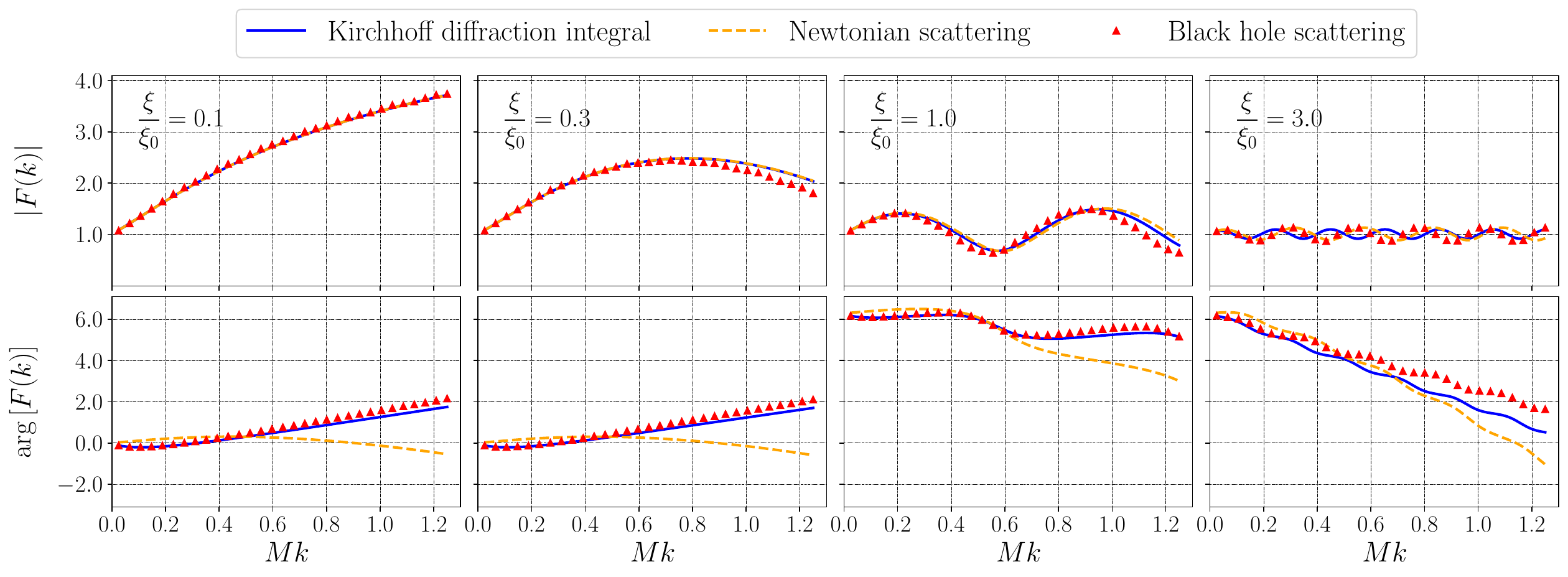}
    \caption{The comparison between three different approaches to calculate the transmission factor $F(k)$, including the Kirchhoff diffraction integral (\ref{eq-4-Kirchhoff}), Newtonian scattering (\ref{eq-4:transmission-factor-Newton}), and BH scattering (\ref{eq-4:transmission-factor-RW}).}
    \label{fig:transmission-factor-comparison}
\end{figure}
\end{widetext}

Such a transmission factor can also be obtained from the scattering model. The BH scattering theory adopts only the linear-perturbation approximation, providing a more rigorous computational approach. The result is given by
\begin{equation}
\label{eq-4:transmission-factor-RW}
\begin{aligned}
F(k)&=e^{-ikr\cos\theta}\tilde{\psi}(k,\bm{r})\\
&=e^{-ikr\cos\theta}\times\,\text{Eq.\,(\ref{eq-4:scattering-wave-function})}.
\end{aligned}
\end{equation}
As the weak-field limit, the Newtonian scattering model provides an analytical result, which can be read off from Eq.\,(\ref{eq-3:scattering-wave-function-paraboloidal}),
\begin{equation}
\label{eq-4:transmission-factor-Newton}
F(k)=e^{-\pi\gamma/2}\Gamma(1+i\gamma){_1F_1}\Big\{-i\gamma,1;2ikr\sin^2(\theta/2)\Big\}.
\end{equation}
To make a comparison between these three results, (\ref{eq-4-Kirchhoff}), (\ref{eq-4:transmission-factor-RW}), and (\ref{eq-4:transmission-factor-Newton}), We set the radius $r=60.0M$ and the angular coordinate as $\xi/\xi_0=\{0.1,0.3,1.0,3.0\}$, which is consistent with the setting in Ref.\,\cite{Takahashi_2003}. The corresponding $\theta$ are listed in Table \ref{tab:parameter-setting}. Performing numerical calculations, the frequency dependence of the above three different transmission factors and their comparison are shown in Fig.\,\ref{fig:transmission-factor-comparison}.

\begin{table}[htbp]
\centering
\renewcommand{\arraystretch}{1.8}
\begin{tabular}{c|c|c|c|c}
\hline\hline
\makebox[1.8cm][c]{$\xi/\xi_0$} & \makebox[1.4cm][c]{0.1} & \makebox[1.4cm][c]{0.3} & \makebox[1.4cm][c]{1.0} & \makebox[1.4cm][c]{3.0} \\
\hline
\makebox[1.8cm][c]{$\theta$ (rad)} & \makebox[1.4cm][c]{0.0258} & \makebox[1.4cm][c]{0.0773} & \makebox[1.4cm][c]{0.2527} & \makebox[1.4cm][c]{0.6591} \\
\hline
\makebox[1.8cm][c]{$\theta$ (deg)} & \makebox[1.4cm][c]{1.4790} & \makebox[1.4cm][c]{4.4293} & \makebox[1.4cm][c]{14.4775} & \makebox[1.4cm][c]{37.7612} \\
\hline\hline
\end{tabular}
\caption{The parameter setting in the numerical calculation of the transmission factor.}
\label{tab:parameter-setting}
\end{table}

The amplitude of the transmission factor, $|F(k)|$, represents the amplification of the lensing signal compared to the unlensed signal. As $\xi/\xi_0$ increases, the deviation between the Newtonian scattering and diffraction integral becomes more pronounced, because Eq.\,(\ref{eq-4-Kirchhoff}) includes a near-axis approximation. Compared to BH scattering, the other two results exhibit greater deviation when $\xi/\xi_0$ is larger. This deviation also becomes slightly more pronounced as frequency increases. The angle of $F(k)$, representing the relative phase delay of the lensed waveform, presents significant differences between these three approaches. These differences stem from that the long-range characteristic of gravity is neglected in the diffraction integral, and the Newtonian and BH scattering adopt different tortoise coordinates.

\section{\label{sec:summary}Summary And Discussion}
Since the establishment of the BH perturbation theory, the BH scattering problem has attracted a lot of attention. Especially in the era of GW astronomy, the BH scattering is more deeply investigated to construct accurate lensed gravitational waveforms and to understand the lensing phenomenon at the wave-optics level. However, the partial-wave method, the main tool for BH scattering, encountered two computational challenges in the previous studies: the PWS divergence and the Poisson-spot divergence. This work aims to explore the cause and response. 

In Section \ref{sec:inc}, we investigate the convergence of PWS for the incident plane wave. Without the asymptotic expansion, Eq.\,(\ref{eq-2:plane-wave-spherical-harmonics-expansion}) is convergent for arbitrary finite radius $r$, and the natural truncation is $\ell_{\max}\sim kr$ (see Fig.\,\ref{fig:plane_wave_convergence}). However, because the large-$kr$ expansion is incorrectly applied to $j_{\ell}(kr)$ with $\ell\sim kr$, the right-hand side of Eq.\,(\ref{eq-2:plane-wave-spherical-harmonics-expansion-asymptotic}) exhibits divergence. The same scheme also appears in the DCS calculation. In the derivation, the large-$kr$ expansion is applied to the radial function $\tilde{R}_{\ell}(k,r)$. However, as shown in Fig.\,\ref{fig:Kummer_asymptotic_expansione} and Fig.\,\ref{fig:RW_asymptotic_expansione}, such expansions are only valid for the region where $kr\gg\ell$ rather than $kr\gg1$. Therefore, the low-$\ell$ (e.g., $\ell\ll kr$) modes is well approximated, while the high-$\ell$ (e.g., $\ell\gtrsim kr$) modes are usually not. This provides a mathematical explanation for the PWS divergence.

Meanwhile, we also provided a physical interpretation of the above discussion, which has been visualized in Fig.\,\ref{fig:Newton-potental} and Fig.\,\ref{fig:RW-potental}. We suppose that the observer is located at the radius $r$, and we denote the frequency of the incident wave by $k$. The observer is far away from low-$\ell$ ($\ell\ll kr$) potential barriers, and the corresponding partial waves contribute to the scattered wave through their incident and reflected parts. In contrast, the high-$\ell$ ($\ell\gtrsim kr$) partial waves at the observer's location are suppressed by a high potential barrier, and only their negligible transmission parts contribute. This is precisely the reason why there is a natural truncation, $\ell_{\max}\sim kr$, in the PWS. However, the large-$kr$ expansion approximates all $\ell$-modes by significant non-zero values, as one assumes that the observer is always far from the turning point for an arbitrary mode. 

In previous studies on BH scattering, the asymptotic expansion is applied to the radial function, which is subsequently used to approximate the scattered wave function by Eq.\,(\ref{eq-3:wave-function-PW-asymptotic-expansion-reduced}). The divergent $f(\theta)$ is regularized to $\widehat{f}(\theta)$ by SRM or another method. However, this regularization is effective only in the far-axis region, and the Poisson-spot divergence still emerges. This is because the harmonics decomposition is pushed to approximate the forward-scattering singularity in the exact wave function. One can realize this from the paraboloidal-coordinate solution of Newtonian scattering. As discussed in subsection \ref{subsec:3-D}, the inappropriate asymptotic expansion, invalid for $2kr\sin^2(\theta/2)\lesssim1$, in deriving the Rutherford formula (\ref{eq-3:Rutherford-formula}) results in a singularity of $f(\theta)$ (\ref{eq-3:angular-factor}). In the framework of partial-wave analysis, the PWS (\ref{eq-3:wave-function-PW-asymptotic-expansion-scattered-wave}) is required to express such a singular function even at its singular point. The invalidity of regularization at the on-axis region emphasizes the essentiality of modeling lensing to compute the scattered waveform at a finite radius, avoiding the asymptotic expansion of the radial function.

In Section \ref{sec:Newton} and Section \ref{sec:RW}, we presented the rigorous calculations on Newtonian scattering and BH scattering by avoiding the asymptotic expansion. The diffraction patterns of the scattered waves are shown in Fig.\,\ref{fig:scattering-wave-Newton} and Fig.\,\ref{fig:scattering-wave-RW}, respectively. As shown in Fig.\,\ref{fig:Newton-consistency} and Fig.\,\ref{fig:RW-convergence}, the involved PWS are convergent for both on-axis and off-axis observers. Subsequently, the waveforms given by our calculations and Eq.\,(\ref{eq-3:wave-function-PW-asymptotic-expansion-reduced}) are compared in Fig.\,\ref{fig:Newton-waveform-angular-distribution} and Fig.\,\ref{fig:RW-waveform-angular-distribution}, showing that Eq.\,(\ref{eq-3:wave-function-PW-asymptotic-expansion-reduced}) fails to approximate the exact wavefunction in the region near the axis. In subsection \ref{subsec:transmission}, we list three different approaches to find the transmission factor for the lensed scalar and gravitational waveforms, and show their comparison in Fig.\,\ref{fig:transmission-factor-comparison}. 

It is beneficial to summarize the computational procedure. Firstly, observers are always located at the finite radius $r$, and PWS are subsequently truncated at $\ell_{\max}\sim kr$. Secondly, the radial equation (e.g., the Coulomb wave equation or the spin-0 RW equation) with $\ell\lesssim\ell_{\max}$ is numerically solved to the outer boundary, denoted by $r_{\max}$, which is required to be sufficiently distant from the turning point. Thirdly, the wave function $\tilde{R}_{\ell}(k,r)$ with $\ell\lesssim\ell_{\max}$ is asymptotically expanded at $r_{\max}$ and matched to the boundary condition. Finally, performing the resummation of PWS gives the full scattered wave function. Compared with the conventional one, our approach first finds $\ell_{\max}$ and then calculates the partial waves with $\ell\lesssim\ell_{\max}$ rather than solving all the partial waves and then finding a suitable truncation.

In our future studies, three significant extensions will be considered. Firstly, this work will be extended to the general scattering model with finite $r_s$ and $r$, with the truncation being $\ell_{\max}\sim k\times\min\{r,r_s\}$ \cite{Kubota_2024}, in which the scattering of spherical waves will be computed. Secondly, the numerical results of this work are limited to the range of $r\lesssim60M$. Although this meets the requirement of $r\gg M$, it is much smaller than the actual astrophysical scale. The large $r$ means that the PWS should be truncated at a large $\ell_{\max}$. As an example, let us consider a planar GW scattered by the Sun, with characteristic frequency $k\sim M_{\odot}^{-1}$, where $M_{\odot}$ is the solar mass. For the detector on Earth, the distance $r$ is approximately $5\times10^7M_{\odot}$, and the PWS truncation is approximately $\ell_{\max}\sim5\times10^7$. However, the extremely large $\ell$ corresponds to an extremely high potential barrier, bringing greater numerical jumps to the wave function near the peak and leaving us a non-negligible numerical challenge. Our future work plans to extend the computation to larger $r$ values and characterize the scattering process at a larger scale, providing a foundation for constructing ready-to-use lensing waveform templates. Finally, GW scattering will be considered, which benefits us in gaining a more comprehensive understanding of the spin-2 nature. A series of related questions, such as gauge transformations, polarization distortions, and helicity conservation, is planned to be carefully investigated.

\begin{acknowledgments}
We thank Jianhua He, Xian Chen, Chengjiang Yin, Yucheng Yin, and the anonymous referee for their helpful discussions. This work is supported by the National Key R\&D Program of China (Grant No. 2022YFC2204602 and 2021YFC2203102) and the National Natural Science Foundation of China (Grant No. 12325301 and 12273035).
\end{acknowledgments}

~

\appendix
\section{\label{app-A}Series reduction method and differential cross section}
SRM aims to accelerate the convergence of PWS involved in $f(\theta)$ (\ref{eq-3:wave-function-PW-asymptotic-expansion-scattered-wave}). For convenience, we write such a series as
\begin{equation}
\label{eq-app-f-theta}
f(\theta)=\sum_{\ell=0}^{\infty}a_{\ell}{\rm P}_{\ell}(\cos\theta),
\end{equation}
where $a_{\ell}$ is
\begin{equation}
a_{\ell}=\frac{(2\ell+1)}{2ik}\Big\{e^{2i\delta_{\ell}}-1\Big\}.
\end{equation}
The $n$-th reduced series is defined as 
\begin{equation}
\begin{aligned}
f^{(n)}(\theta)
&=(1-\cos\theta)^{n}
\sum_{\ell=0}^{\infty}a_{\ell}{\rm P}_{\ell}(\cos\theta)\\
&=\sum_{\ell=0}^{\infty}a^{(n)}_{\ell}{\rm P}_{\ell}(\cos\theta).
\end{aligned}
\end{equation}
The coefficients $a^{(n)}_{\ell}$ satisfying the following recurrence relation,
\begin{equation}
a^{(n+1)}_{\ell}
=a^{(n)}_{\ell}
-\frac{\ell}{2\ell-1}a^{(n)}_{\ell-1}
-\frac{\ell+1}{2\ell+3}a^{(n)}_{\ell+1},
\end{equation}
where we usually set $a^{(n)}_{-1}=0$. The series $f(\theta)$ is finally reduced as 
\begin{equation}
\widehat{f}(\theta)=(1-\cos\theta)^{-n}f^{(n)}(\theta).
\end{equation}

The original coefficient $a_{\ell}$ and the first three orders of reduced coefficients, $a_{\ell}^{(n)}, (n=1,2,3)$, are plotted in Fig.\,\ref{fig:series-reduction}. While the original coefficients show no sign of convergence in both Newtonian and BH scattering, the reduced coefficients decrease rapidly as $\ell$ increases, confirming that SRM has successfully improved the convergence of $f(\theta)$. Using $a_{\ell}$ and $a_{\ell}^{(n)}, (n=1,2,3)$, we calculate the DCS and present their numerical results in Fig.\,\ref{fig:differential-cross-section}. In Newtonian scattering, the SRM performs well but the on-axis divergence is still left. For BH scattering, the SRM remains effective only within the far-axis region, with $\theta>0.1\pi$, while also leaving divergence in the near-axis region.

\begin{widetext}

\begin{figure}[H]
    \centering
    \includegraphics[width=1.0\linewidth]{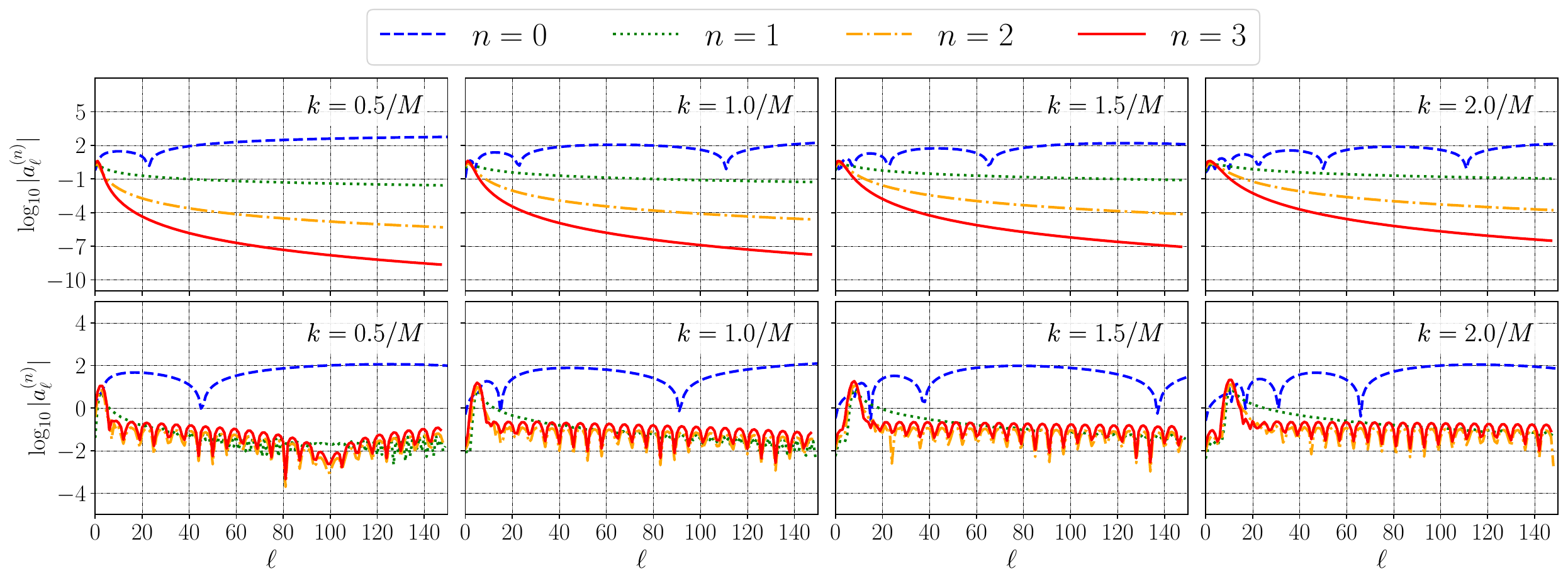}
    \caption{Original coefficients $a_{\ell}$ and the first three orders of reduced coefficients $a_{\ell}^{(n)}\,(n=1,2,3)$ as functions of $\ell$. The reduced coefficients exhibit significantly improved convergence properties. The first row presents results for Newtonian scattering, while the second row corresponds to BH scattering calculations.}
    \label{fig:series-reduction}
\end{figure}
\begin{figure}[H]
    \centering
    \includegraphics[width=1.0\linewidth]{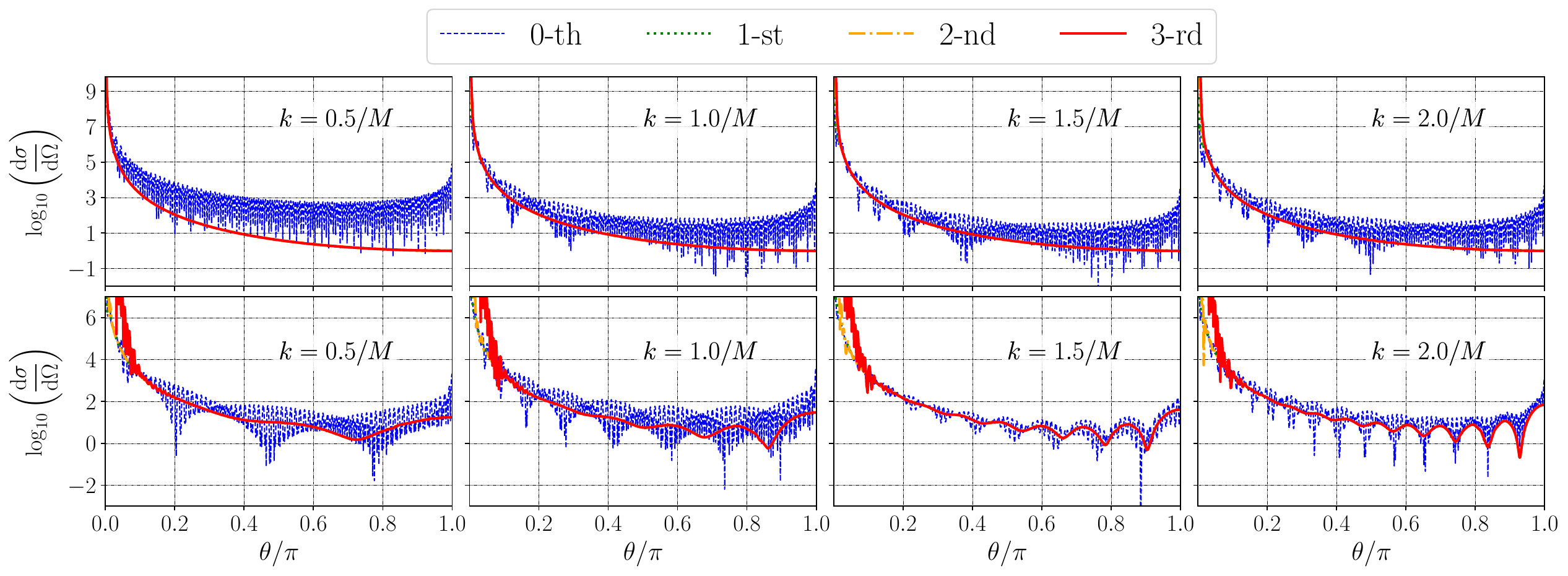}
    \caption{DCS computed from $a_{\ell}$ and $a_{\ell}^{(n)}\,(n=1,2,3)$. The first row presents results for Newtonian scattering, while the second row shows BH scattering calculations. In Newtonian scattering, the DCS converges for all $\theta$ values except for the optical axis. For BH scattering, the DCS exhibits poor convergence in the near-axis region, accompanied by a divergence behavior.}
    \label{fig:differential-cross-section}
\end{figure}
\end{widetext}

\section{\label{app-B}Eikonal expansion of linearized Einstein equation}
Based on the vacuum Einstein field equation $R_{\mu\nu}=0$, one decomposes the full metric $g_{\mu\nu}$ into background and GWs, $g_{\mu\nu}=\gamma_{\mu\nu}+h_{\mu\nu}$, and then derives the LEFE, given by
\begin{equation}\label{LEFE}
\Box^2\psi_{\mu\nu}(x)+2R^{\rm(B)}_{\alpha\nu\beta\mu}\psi^{\alpha\beta}(x)=0,
\end{equation}
where $R^{\rm(B)}_{\alpha\nu\beta\mu}$ is the Riemann tensor of background. The trace-reverse metric $\psi_{\mu\nu}$ is defined as 
\begin{equation}
\psi_{\mu\nu}\equiv h_{\mu\nu}-\frac{1}{2}\gamma_{\mu\nu}\gamma^{\alpha\beta}h_{\alpha\beta},
\end{equation}
satisfying the Lorentz gauge $\nabla_{\mu}\psi^{\mu\nu}=0$ and traceless condition $\gamma_{\mu\nu}\psi^{\mu\nu}=0$.

Expanding the GW metric in terms of the characteristic frequency $k$, we have
\begin{equation}
\label{eq-app-eikonal-expansion-GW}
\psi_{\mu\nu}(x)={\rm Re}\,\Big\{\bar{\psi}(x) e_{\mu\nu}e^{ik\Phi(x)}\Big\},
\end{equation}
where $k$ is the GW frequency, satisfying $k\mathcal{R}\gg1$, with $\mathcal{R}$ being the characteristic scale of background. $\Phi$ is the GW phase, and its gradient is defined as the null wavevector. i.e., $q_{\mu}=-\partial_{\mu}\Phi(x)$. The polarization tensor $e_{\mu\nu}$ is required to be transversal and parallel-transported, i.e., $q^{\mu}e_{\mu\nu}=0$, and $q^{\alpha}\nabla_{\alpha}e_{\mu\nu}=0$. Substituting the expansion (\ref{eq-app-eikonal-expansion-GW}) into the LEFE (\ref{LEFE}), one gets the evolution equation of amplitude $\bar{\psi}(x)$ at $\mathcal{O}(k)$ order, which is
\begin{equation}
\label{eq-app:evolution-equation-amplitude}
q_{\alpha}\nabla^{\alpha}\bar{\psi}+\frac{1}{2}(\nabla_{\alpha}q^{\alpha})\bar{\psi}=0.
\end{equation}
Meanwhile, since $q_{\mu}$ is the gradient of a scalar function, it satisfies $\nabla_{\mu}q_{\nu}=\nabla_{\nu}q_{\mu}$. Using this property, together with the null condition $q_{\alpha}q^{\alpha}=0$, one derives the geodesic equation for $q_{\mu}$, $q^{\alpha}\nabla_{\alpha}q_{\mu}=0$ \cite{Andersson2021}. Subsequently, the dynamics of high-frequency GWs are governed by the null-geodesic behavior. This is the emergence of the equivalence principle of general relativity in the wave-scattering regime.

Similar to Eq.\,(\ref{eq-app-eikonal-expansion-GW}), performing an analogous expansion for the KG equation (\ref{eq-3:KG}) gives
\begin{equation}
\label{eq-app-eikonal-expansion-scalar}
\psi(x)={\rm Re}\,\Big\{\bar{\psi}(x)e^{ik\Phi(x)}\Big\}
\end{equation}
and reveals that the amplitude of scalar waves likewise satisfies Eq.\,(\ref{eq-app:evolution-equation-amplitude}). This indicates that the scalar amplitude of high-frequency GWs and scalar waves exhibits the same behavior. This consequently establishes the foundation for employing scalar fields to approximate high-frequency GWs.

\bibliographystyle{apsrev4-2}
\bibliography{ref.bib}

\end{document}